\documentclass[prd,nofootinbib,preprint,superscriptaddress]{revtex4-1}

\usepackage{amsmath, amssymb, amsthm, graphicx, epsfig, fancyhdr,epsfig}
\usepackage{physics}
\usepackage{ulem}

\usepackage{tikz-feynman}
\tikzfeynmanset{compat=1.1.0}

\usepackage{tikzsymbols}
\usepackage{amsmath,amssymb}
\usepackage{tikz}
\usepackage{tikz-feynman}
\usetikzlibrary{calc,positioning,shapes.misc}
\tikzfeynmanset{compat=1.1.0}

\tikzfeynmanset{
 every fermion/.style={/tikz/line width=0.6pt},
 every scalar/.style={
  /tikz/dashed,
  /tikz/dash pattern=on 3pt off 3pt,
  /tikz/line width=0.6pt
 }
}
\tikzset{
 massX/.style={
  cross out,
  draw=black,
  line width=0.6pt,
  minimum size=5.5pt,
  inner sep=0pt
 }
}

\usepackage{float}
\usepackage{xcolor}

\usepackage[caption=false]{subfig}

\newcommand{\be}{\begin{equation}}
\newcommand{\ee}{\end{equation}}
\newcommand{\bea}{\begin{eqnarray}}
\newcommand{\eea}{\end{eqnarray}}

\begin{document}
\title{Gravitational Waves from hybrid defects as probe of Flavor symmetry breaking: \it{Machine-Learning Approach}}

\author{Anish Ghoshal}
\email{a.ghoshal@sussex.ac.uk}
\affiliation{Department of Physics and Astronomy, University of Sussex, \\
Brighton, BN1 9RH, United Kingdom}

\author{Ilia Gogoladze}
\email{iliag@udel.edu}
\affiliation{Department of Physics and Astronomy, University of Delaware, \\ Newark, DE 19716, USA}

\author{Amit Tiwari}
\email{amitiit@mit.edu}
\affiliation{Massachusetts Institute of Technology, Cambridge, MA 02139, USA}

\begin{abstract}
\textit{We present a novel possibility that a network of domain walls bounded by cosmic strings generates a stochastic gravitational wave background (SWGB) signal originating from the spontaneous breaking of a gauged $U(1)_F$ flavor symmetry and the subsequent breaking of discrete $Z_2$ symmetry that accommodates dark matter. The gravitational wave (GW) spectrum produced by the string-bounded-wall network can be detected for high $U(1)_F$ breaking scales in forthcoming   GW detectors including LISA, ET and SKA. The GW signal exhibits a distinctive frequency slope, in the infrared, compared to the standard cosmic-string case, in the frequency range between micro-hertz and hertz.
We develop a possible strategy to distinguish and characterize GW spectrum of the hybrid defect from from other defects, such as stable cosmic strings, via employing the exact calculation with a \textit{machine-learning surrogate} based on a multilayer perceptron (MLP), trained on spectra obtained from the full numerical treatment. This is then used for rapid inference in the detector-specific  signal-to-noise ratio (SNR) computation which also makes the process fast and efficient. We also discuss some possible complementarity between GW searches and Flavor observables in the laboratory.
}
\end{abstract}

\maketitle

\tableofcontents

\section{Introduction}
 
The Standard Model (SM) provides a robust framework for particle interactions up to the TeV scale, yet it fails to explain the pronounced hierarchies within the fermion sector. One widely studied approach to this problem involves extending the SM with flavor-dependent gauge symmetries that generate hierarchical Yukawa couplings via spontaneous symmetry breaking (SSB). The Froggatt-Nielsen (FN) mechanism \cite{Froggatt:1978nt, Berezhiani:1983hm, Dimopoulos:1983rz} represents a leading framework in this category, employing a
$U(1)_F$ symmetry to translate small charge differences into the large observed mass variations. Under this framework, all underlying Yukawa couplings are assumed to be $\mathcal{O}(1)$. The effective couplings for light fermions are generated via higher-dimensional, non-renormalizable operators, resulting in a suppression by powers of the expansion parameter $\epsilon \equiv \langle \Phi \rangle/\Lambda$. In this expression, $\langle \Phi \rangle$ denotes the vacuum expectation value of the flavon ($\Phi$) field that spontaneously breaks $U(1)_F$ flavor symmetry, while $\Lambda$ represents the mass scale of the integrated-out messenger sector. In this approach, the vast hierarchy between fermion masses exemplified by the up-to-top quark mass ratio of approximately $10^{-6}$ is successfully generated using $U(1)_F$ flavor charges that differ only by $\mathcal{O}(1)$ factors. The $U(1)_F$ flavor symmetry is broken to a discrete $Z_2$ subgroup by introducing a complex scalar field $\Phi$ with a flavor charge $q_{\Phi}=2$. Since the first homotopy group of the vacuum manifold $U(1)/Z_2$ is non-trivial, with $\pi_1(U(1)/Z_2) \simeq Z$, this produces cosmic strings. 

The Yukawa couplings are suppressed due to the presence of a complex scalar singlet field $S$ with a $U(1)_F$ charge $q_{S}=1$. Once the $S$ field develops a vacuum expectation value (VEV), the remaining $Z_2$ symmetry is also broken. For the second stage of symmetry breaking, $Z_2 \rightarrow 1$, the zeroth homotopy group of the vacuum manifold $(Z_2/1)$ is non-trivial because it is disconnected, with $\pi_0(Z_2/1) \cong Z_2$. 
Note that after the CP-even real scalar component of the $S$ field develops a VEV, the CP-odd imaginary scalar component of the $S$ field, $a$, can still retain a residual $Z_2$ symmetry (or sometimes called dark CP symmetry). Since the SM fields are even under this $Z_2$ symmetry and the pseudo-scalar $a$ field cannot mix with real scalars, it cannot decay into any SM final states. This makes the pseudo-scalar $a$ field an excellent candidate for dark matter (DM) \cite{Barger:2008jx, Gonderinger:2012rd, Gross:2017dan}. Due to this $Z_2 \rightarrow \mathcal{I}$, we show that the flavor model along with DM naturally accommodates a hybrid topological defect containing domain walls bounded by local cosmic strings or simply, \textit{walls-bounded-by-strings}.

 In general, the spontaneous breaking of global or local symmetries typically leads to the formation of both topological and non-topological defects \cite{Kibble:1976sj, Kibble:1980mv}. Within the specific framework of the $U(1)_F$ gauge symmetry, two distinct topological configurations may arise: stable cosmic strings and domain walls bounded by strings \cite{Kibble:1982dd,Everett:1982nm}. The former case occurs when the $U(1)_F$ symmetry is spontaneously broken to the identity. The latter scenario is realized through a two-stage symmetry-breaking pattern, wherein $U(1)_F$ is first broken to a discrete $Z_N$
subgroup, generating a network of cosmic strings. A subsequent breaking of the $Z_N$ symmetry then leads to the formation of domain walls, with the pre-existing strings serving as their boundaries. This hybrid topological defect, commonly referred to as a "walls bounded by strings" configuration \cite{Kibble:1982dd,Everett:1982nm,Lazarides:2023iim}, consists of cosmic strings connected to domain walls and is recognized as a possible source of stochastic gravitational wave backgrounds (SGWB) \cite{Dunsky:2021tih, Maji:2023fba, Lazarides:2023ksx, Fu:2024rsm}. In the context of early-universe cosmology, the evolution and subsequent decay of such a network can provide distinct signatures detectable by current and future gravitational wave (GW) interferometers; see e.g. \cite{Janssen:2014dka, Sesana:2019vho, LISA:2017pwj, Kudoh:2005as, Kawamura:2020pcg, Harry:2006fi, AEDGE:2019nxb, Badurina:2019hst, LIGOScientific:2016wof, Hild:2008ng, LIGOScientific:2022sts, KAGRA:2021kbb, Jiang:2022uxp, NANOGrav:2023gor, NANOGrav:2023hvm,EPTA:2023sfo, EPTA:2023fyk}. There are several studies on the possible origins of SGWB involving continuous flavor $U(1)_F$ symmetry. For instance, the stable cosmic strings associated with the spontaneous breaking of the gauge $U(1)_F$ symmetry have been discussed in light of flavor observables and SGWB in \cite{Blasi:2024vew} relying on paticle production from strings, and for SU(2) case, in Ref. \cite{Antusch:2025xrs} which also involves an additional inflationary set up \footnote{First-order phase transitions as sources of SGWB in the framework of the $U(1)_F$ flavor symmetry have received considerable attention; see, e.g., \cite{Baldes:2016gaf, Greljo:2019xan, Ringe:2022rjx}.}. We do not invoke any non-minimal features apart in our scenario. 

Since a detailed assessment of the detectability of the hybrid GW signal requires repeated evaluation of the spectrum over a wide region of the parameter space and across the frequency bands of different experiments, a direct numerical signal-to-noise ratio (SNR) analysis becomes computationally expensive. The GW signal exhibits a distinctive frequency slope, in the infrared, compared to the standard cosmic-string case, in the frequency range between micro-hertz and hertz. We  develop a strategy such that the characteristic GW spectrum from the hybrid defect may be analyzed to distinguish it from other defects, such as stable cosmic strings, under some assumptions. To quantify whether a given detector band actually sees the hybrid spectrum portion of the signal, we define a turning-point frequency \(f_{\rm turn}\) in the GW spectrum, by directly comparing the hybrid spectrum to the pure-string spectrum at fixed $U(1)_{F}$ flavor breaking scale \(v_F\). In order to employ this strategy, we complement the exact calculation with a \textit{machine-learning surrogate} based on a multilayer perceptron (MLP), trained on spectra obtained from the full numerical treatment and then used for rapid inference in the SNR computation. This allows us to efficiently map the detectability of the hybrid signal and quantify the region in which it can be distinguished from the corresponding pure cosmic-string spectrum.

\textit{The paper is structured as follows:} in Section II, we describe the flavor model; in Section III, we briefly discuss the dark matter candidate that naturally arises in our model; and in Section IV, we discuss the origin of the walls-bounded-by-strings topological defect. 
The gravitational waves emitted by the formation of topological defects involving such flavor symmetry breaking are presented in Section V. In Sections VI and VII, we present the SGWB phenomenology and the machine-learning-assisted SNR analysis. Our conclusions are given in Section VIII.


\section{ Model with $U(1)_F$ Flavor symmetry }
Understanding the fermion mass hierarchy remains an unsolved problem within the Standard Model (SM) of particle physics. A key challenge involves explaining why the observed mass ratios of quarks and charged leptons align with the order-of-magnitude estimations 
\cite{ParticleDataGroup:2024cfk}
\begin{align}
m_t:m_c:m_u \approx 1: 10^{-2} : 10^{-5}, \nonumber \\
m_b : m_s: m_d \approx 1: 10^{-2} : 10^{-3}, \nonumber \\
m_{\tau}: m_{\mu} : m_{e} \approx 1:10^{-1}:10^{-3},
\label{m-hier}
\end{align}
where $m$ represents the masses; the subscripts $t$, $c$, $u$, $b$, $s$, and $d$ represent the top, charm, up, bottom, strange, and down quarks; and $\tau$, $\mu$, and $e$ denote the tau, muon, and electron masses.

In this section, we develop a model to explain dynamically such fermion mass hierarchies and mixing by implementing the Froggatt-Nielsen (FN) mechanism \cite{Froggatt:1978nt} which is not accommodated within the SM. The basic idea is to use a gauged (or global) Abelian flavor $U(1)_F$ symmetry to forbid all tree-level Yukawa couplings except for the top quark.
Fermion mass and mixing hierarchies are then generated via higher-dimensional operators involving an SM-singlet scalar field ($\Phi$). The field $\Phi$ develops a vacuum expectation value (VEV) that breaks the $U(1)_{F}$ symmetry at some high scale and results in the SM Yukawa couplings. 
In this approach, all the Standard Model (SM) fermions carry $U(1)_F$ charges ${q}_{f_i}$. Here, $f_i$ denotes the SM fermions, and $i= 1,\,2,\,3$ is the generation index. We use the standard notation for the SM fermion fields: $Q_i$ and $L_i$ are left-handed quark and lepton doublets, $u_i$ and $d_i$ are quark singlets, and $e_i$ and $\nu_i$ are the right-handed electron and right-handed singlet neutrino fields, respectively; $i$ stands for the family index. In the scalar sector, we have the following fields: $H$, the SM Higgs, and the flavons ($\Phi$ and $S$), which are complex scalar fields and SM singlets but have non-trivial $U(1)_F$ charges. For a more detailed description of particle content in our model, see Table~\ref{tcharge}. \footnote{$Z_2$ symmetry charges can be written using '+' and '-' notation \cite{Fu:2024rsm}.}
\begin{table}[h]
\begin{center}
\begin{tabular}{|c|c|c|c|}\hline
\rule[5mm]{0mm}{0pt}Field & $SU(3)_C\times SU(2)_L\times U(1)_Y$ & $U(1)_F$ & \, $Z_2\subset U(1)_F$  \,\\\hline
\rule[5mm]{0mm}{0pt} \,$Q_1$, $Q_2$, $Q_3$ \,& $(3,\, 2,\, 1/6)$ & \,\,$ 8, \,4, \,0$ \, & \, $ 0, 0, 0 $ \, \\
\rule[5mm]{0mm}{0pt}$u_1$, $u_2$, $u_3$ & $(3,\, 1,\, 2/3)$ & $ -8, \, -4,\, 0$ \, & $ \, 0, 0, 0 \,$ \\
\rule[5mm]{0mm}{0pt}$d_1$, $d_2$, $d_3$ & $(3,\, 1,\, -1/3)$ & $ -8,\, -6,\,
6$ \, & $ \, 0, 0, 0 \,$ \\
\rule[5mm]{0mm}{0pt}$L_1$, $L_2$, $L_3$& $(1,\, 2,\, -1/2)$ &\,\,$ 2+2p,\, 2p,
\,2p $ & $ \, 0, 0, 0 \,$ \\
\rule[5mm]{0mm}{0pt}$e_1$, $e_2$, $e_3$ & (1,\, 1,\, -1)$ $&\,\,$ -14+2p,\, -10+2p,\, -6+2p$\,& $ \, 0, 0, 0 \,$ \\
\rule[5mm]{0mm}{0pt}$\nu_1$, $\nu_2$,
$\nu_3$ & ( 1,\, 1,\, 0) & $ -2,\, 0,\, 0$ & $ \, 0, 0, 0 \,$ \\
\rule[5mm]{0mm}{0pt}$H$ & $(1,\, 2,\, 1/2)$ & $ 0 $ & $ \, 0 \,$ \\
\rule[5mm]{0mm}{0pt} $\Phi$, $ S$ & (1,\, 1,\, 0) & $ 2,\, 1 $ & $ \, 0, 1 \,$ \\
\hline
\end{tabular}
\caption{\it Gauge charge assignments for
the SM fields and the flavon fields $\Phi$ and $S$. The subscripts (1, 2, 3) stand for family indices. Here $p$ is an arbitrary number. The positive integer $p$ provides the degrees of freedom necessary to reproduce the observed neutrino mixing and mass differences (see Eq.~(7)), for a fixed right-handed neutrino mass scale.}
 \label{tcharge}
\end{center}
\end{table}
The general Yukawa sector of the SM Lagrangian which is also invariant under the $U(1)_F$ symmetry has the following expression
\begin{align}
-\mathcal{L}_{\rm Yuk} =
Y_{ij}^u \, {\overline{Q}}_i u_j\, \tilde{H} \left( \frac{\Phi}{\Lambda} \right)^{(n_1)_{ij}} + \,
Y_{ij}^d \, {\overline{Q}}_i d_j\, H \, \left( \frac{\Phi}{\Lambda} \right)^{(n_2)_{ij}} + \, Y_{ij}^\ell \, {\overline{L}}_i e_j\, H \,\left( \frac{\Phi}{\Lambda} \right)^{(n_3)_{ij}} + \nonumber \\
Y_{ij}^{\nu_D} \, {\overline{L}}_i \nu_j\, \tilde{H} \left( \frac{\Phi} {\Lambda} \right)^{(n_4)_{ij}} + \, M_{R_{ij}} {\overline{\nu}^c}_i \nu_j\, \left( \frac{\Phi}{\Lambda} \right)^{(n_5)_{ij}} + {\rm h.c.} \,
\label{eq:Yuk1}
\end{align}
Here, $\tilde{H}=-i\sigma_2 H^*$ is complex conjugate scalar field. 
$M_{R_{ij}}$ is arbitrary mass term. 
All non-renormalizable couplings in Eq.~(\ref{eq:Yuk1}) are obtained after integrating out the vector-like (non-chiral) fermions ($\Psi_i+ \overline{\Psi}_i$). $i=1\, ... \,n$ where $n$ correspond to the numerical values of power of non-renormalizable couplings given in Eq.~(\ref{eq:Yuk1}).
For simplicity, we assume all vector like particles have the mass $M_i\overline{\Psi}_i\Psi_i$ around $\Lambda$ scale and adapt $M_i \equiv \Lambda$. 

The $U(1)_F$ gauge invariance condition for each Yukawa coupling in Eq.~(\ref{eq:Yuk1}) is given by
\begin{align}
\Sigma q_{\alpha}=0
\label{gauge}
\end{align}
where $q_{\alpha}$ represents the charge of the fields involved in each coupling. For example, the gauge invariance condition for down-type quarks is:
\begin{align}
-q_{Q_i}+q_{d_j}+q_H+n_{ij}q_{\Phi}=0
\label{gauge-5}
\end{align}
determine the $n_{ij}$  values, which are expected to be positive integers. Using the gauge invariance condition in Eq.~(\ref{gauge}) for all Yukawa couplings in Eq.~(\ref{eq:Yuk1}), we can express the exponent of each term in the Yukawa sector as a combination of $U(1)_F$ charges
\begin{align}
(n_1)_{ij} \equiv \frac{{q}_{Q_i} -{q}_{u_j} + {q}_{H}}{q_{\Phi}}\,,\quad
(n_2)_{ij} \equiv \frac{{q}_{Q_i} -{q}_{d_j} - {q}_{H}}{q_{\Phi}}\,,\quad 
(n_3)_{ij} \equiv \frac{{q}_{L_i} -{q}_{e_j} - {q}_{H}}{q_{\Phi}}\,, \quad \nonumber \\
(n_4)_{ij} \equiv \frac{{q}_{L_i} -{q}_{\nu_j} + {q}_{H}}{q_{\Phi}}\,,\quad
(n_5)_{ij} \equiv \frac{-{q}_{\nu_i} -{q}_{\nu_j} + {q}_{\Phi}}{q_{\Phi}}\,,\quad
\label{eq:ex1}
\end{align}
As shown above, all parameters in the powers of the Yukawa couplings in Eq.~(\ref{eq:Yuk1}) have an overall factor proportional to the $U(1)_{F}$ charge of the $\Phi$ field. Therefore, in general, the $\Phi$ field can be assigned any $U(1)_{F}$ charge value and still yield the same power for the non-renormalizable couplings in Eq.~(\ref{eq:Yuk1}), provided that all $U(1)_{F}$ charges for the SM fields are multiplied by the same overall factor. Assuming integer values for the $U(1)_{F}$ charges, a choice of $q_{\Phi}\neq 1$ will lead to the breaking of the gauge $U(1)_{F}$ symmetry down to a $Z_{q_{\Phi}}$ discrete symmetry. The $S$ field, with $q_{S} =1$, will completely break this residual discrete symmetry.

For simplicity, we consider the case where $q_{\Phi}=2$.
The $U(1)_F \rightarrow Z_2\rightarrow I$ symmetry-breaking chain leads to the creation of domain walls bounded by strings (a hybrid topological defect) \cite{Kibble:1982dd,Everett:1982nm}. This hybrid defect makes the model testable via the gravitational wave (GW) spectral shapes that we discuss later on.

We can construct a similar Yukawa sector to the one given in Eq.~(\ref{eq:Yuk1}) by incorporating the scalar field $S$. Replacing $q_{\Phi}$ with $q_{S}$ in the gauge invariance condition of Eq.~(\ref{gauge-5}) for down quarks, we obtain
\begin{align}
-q_{Q_i}+q_{d_j}+q_H+ 2n_{ij}q_{S}=0
\label{gauge-7}
\end{align}
Here the factor of 2 appears because the $U(1)_F$ charge of the $\Phi$ field is twice as large as the charge of the $S$ field. This means that for the non-renormalizable Yukawa couplings, instead of $(\phi/\Lambda)^n_{ij}$ suppression factor, we will have 
$(S/\Lambda)^{2n_{ij}}$.
As a result, the non-renormalizable Yukawa couplings in Eq.~(\ref{eq:Yuk1}) involving the $S$ field will be more suppressed than the terms involving only the $\Phi$ field. 
Consequently, all non-renormalizable Yukawa couplings that involve the $S$ field can be neglected in our model.

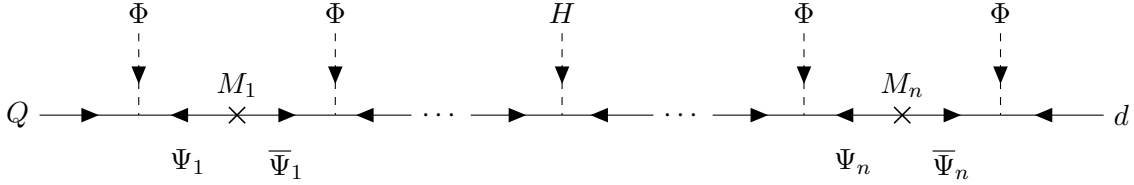
\begin{figure}[h]
\centering
\begin{tikzpicture}
\begin{feynman}
 \vertex (Q) at (0.0,0.0) {$Q$};
 \vertex (d) at (14.6,0.0) {$d$};

 \vertex[coordinate] (v1) at (1.6,0.0) {};
 \vertex[coordinate] (m1) at (2.9,0.0) {};
 \vertex[coordinate] (v2) at (4.2,0.0) {};
 \vertex      (dots1) at (5.6,0.0) {$\cdots$};
 \vertex[coordinate] (vh) at (7.2,0.0) {};
 \vertex      (dots2) at (8.8,0.0) {$\cdots$};
 \vertex[coordinate] (vn1) at (10.4,0.0) {};
 \vertex[coordinate] (mn) at (11.7,0.0) {};
 \vertex[coordinate] (vn2) at (13.0,0.0) {};

 \vertex (phi1) at (1.6,1.35) {$\Phi$};
 \vertex (phi2) at (4.2,1.35) {$\Phi$};
 \vertex (H)  at (7.2,1.35) {$H$};
 \vertex (phi3) at (10.4,1.35) {$\Phi$};
 \vertex (phi4) at (13.0,1.35) {$\Phi$};

 \diagram*{
  (Q) -- [fermion]   (v1)
     -- [anti fermion] (m1)
     -- [fermion]   (v2)
     -- [anti fermion] (dots1)
     -- [fermion]   (vh)
     -- [anti fermion] (dots2)
     -- [fermion]   (vn1)
     -- [anti fermion] (mn)
     -- [fermion]   (vn2)
     -- [anti fermion] (d),

  (phi1) -- [charged scalar] (v1),
  (phi2) -- [charged scalar] (v2),
  (H)  -- [charged scalar] (vh),
  (phi3) -- [charged scalar] (vn1),
  (phi4) -- [charged scalar] (vn2),
 };

 \node[massX] at (m1) {};
 \node[massX] at (mn) {};
 \node[above=3pt] at (m1) {$M_1$};
 \node[above=3pt] at (mn) {$M_n$};

 \node[below=9pt] at ($(v1)!0.5!(m1)$) {$\Psi_1$};
 \node[below=9pt] at ($(m1)!0.5!(v2)$) {$\overline{\Psi}_1$};
 \node[below=9pt] at ($(vn1)!0.5!(mn)$) {$\Psi_n$};
 \node[below=9pt] at ($(mn)!0.5!(vn2)$) {$\overline{\Psi}_n$};
\end{feynman}
\end{tikzpicture}

  \caption{\it Example of a Feynman diagram corresponding to the generation of down-type quark
   mass via the non-renormalizable coupling presented in Eq.~(\ref{eq:Yuk1}). The integer $n$ corresponds to the power of the non-renormalizable coupling in Eq.~(\ref{eq:Yuk1}).
   $\Psi_i$ and $\overline{\Psi}_i$ are vector-like particles with masses of order $\Lambda$. $i = 1, \ldots, n$. Dots in the diagram denote $\Phi$ field insertions, with the power determined by the charge assignments in Table~\ref{tcharge}. }
  \label{diagram-1}
\end{figure}

We assume that all Yukawa couplings in Eq.~(\ref{eq:Yuk1}) are of order one. After the $\Phi$ field develops a VEV, one obtains an effective Yukawa coupling for u-quarks:
\begin{align}
y_{ij}=Y_{ij}^u \, \left( \frac{\left\langle \Phi \right\rangle}{\Lambda} \right)^{(n_1)_{ij}}
\label{Yuk-hier}
\end{align}
If $\epsilon \equiv\left\langle \Phi \right\rangle/{\Lambda}$ is a small number and the
set of $q_{f_i}$ charges is sufficiently diversified, one may reproduce various mass and mixing hierarchies within the theory. As an example, we present in Fig.~\ref{diagram-1} a Feynman diagram corresponding to the generation of down-type quark mass via the non-renormalizable coupling presented in Eq.~(\ref{eq:Yuk1}).
An acceptable flavor texture that gives the correct pattern of fermion masses and mixing (see for instance \cite{Binetruy:1994ru, Binetruy:1996xk, Babu:2002tx,Babu:2003zz}) can be obtained from Eq.~(\ref{matr1}) by using the charge assignment given in Table~\ref{tcharge} and has the following structure:
\begin{eqnarray}
\label{matr1}
 M_U&=&\left(
\begin{array}{ccc}
 \epsilon^{8} & \epsilon^{6} & \epsilon^{4} \\
 \epsilon^{6} & \epsilon^{4} & \epsilon^{2} \\
 \epsilon^{4} & \epsilon^{2} & 1 \\ 
\end{array}
\right)\langle H\rangle,
~~~~ M_D=\left(
\begin{array}{ccc}
 \epsilon^{5} & \epsilon^{4} & \epsilon^{4} \\
 \epsilon^{3} & \epsilon^{2} & \epsilon^{2} \\
 \epsilon^{1} & 1 & 1 \\ 
\end{array}
\right)\epsilon^{3}\langle H\rangle, \nonumber\\ \nonumber \\
M_e &=&\left(
\begin{array}{ccc}
 \epsilon^{5} & \epsilon^{3} & \epsilon \\
 \epsilon^{4} & \epsilon^{2} & 1 \\
 \epsilon^{4} & \epsilon^{2} & 1 \\ 
\end{array}
\right)\epsilon^{3}\langle H\rangle, ~~~~~ M_{\nu_D}=\left(
\begin{array}{ccc}
 \epsilon^{2} & \epsilon & \epsilon \\
 \epsilon & 1 & 1 \\
 \epsilon & 1 & 1 \\ 
\end{array}
\right)\epsilon^{p}\langle H\rangle, \nonumber\\ \nonumber \\
M_{\nu_M}&=&\left(
\begin{array}{ccc}
 \epsilon^{2} & \epsilon & \epsilon \\
 \epsilon & 1 & 1 \\
 \epsilon & 1 & 1 \\
\end{array}
\right)M_M, ~~~~~ M^{lighr}_{\nu}=\left(
\begin{array}{ccc}
 \epsilon^{2} & \epsilon & \epsilon \\
 \epsilon & 1 & 1 \\
 \epsilon & 1 & 1 \\
\end{array}
\right)\frac{\langle H\rangle^2}{M_M}\, \epsilon^{2p}
\end{eqnarray}
here $\left\langle H \right\rangle$ denotes the VEV of the SM Higgs doublet. 
 $M_u$, $M_D$, $M_e$ are the up-quark, down-quark, and charged-lepton mass matrices.  $M_{\nu_D}$ is the Dirac 
neutrino Yukawa matrix, and $M_{\nu_M}$ is the right-handed neutrino Majorana mass matrix. The Dirac 
neutrino Yukawa matrix $M_{\nu_D}$ has the overall factor
$\epsilon^p$, where $p$ is an integer. 
For simplicity we use $M_{M} \equiv M_{M_{R_{ij}}}$, see Eq.~(\ref{eq:Yuk1}).
The undetermined overall factor $\epsilon^p$ is related to the possibility that the neutrino is a Majorana particle. We adopt the type-I seesaw mechanism \cite{Minkowski:1977sc, Yanagida:1979as, Gell-Mann:1979vob, Mohapatra:1979ia}
 as a way to obtain light neutrino masses and mixing
\begin{eqnarray}
 \label{seesaw}
M^{light}_{\nu} = M_{\nu_D}^T \, M_{\nu_M}^{-1} \, M_{\nu_D}
\end{eqnarray}
Since Dirac neutrino masses have not yet been measured experimentally, we cannot fix the overall factor $\epsilon^p$, unlike in the case of quarks and charged leptons. The undetermined values of $p$ and $M_M$ in Eq.~(\ref{matr1}) give additional freedom to obtain the correct neutrino mass. The current experimental bound on the individual neutrino mass is $m_{\nu } < 0.45$ eV \cite{KATRIN:2021uub,KATRIN:2024cdt}, and future sensitivity can reach up to $m_{\nu } < 0.2$ eV \cite{Dolinski:2019nrj}.
We have not exhibited order one coefficients in
the matrix elements of Eq.~(\ref{matr1}). We choose the expansion parameter to be $\epsilon \sim 0.2$. In general, $\epsilon$ can take any value in the interval $0<\epsilon<1$. 

We consider here one specific scenario where the $U(1)_{F}$ symmetry is broken at the $10^{14}$ GeV scale. We perform a numerical fit for the possible coefficients in front of the $\epsilon$ parameters in Eq.~(\ref{matr1}), which result from $ \mathcal{O}(1)$ Yukawa couplings involving the non-renormalizable interactions discussed in Eq.~(\ref{eq:Yuk1}). In our fit, we utilize data provided in \cite{Antusch:2025fpm}, where the authors employed the Particle Data Group (PDG) dataset \cite{ParticleDataGroup:2024cfk} to calculate quark and lepton Yukawa couplings and mixing at different scales using two-loop Renormalization Group Equations (RGE). We use the following values of up- and down-quark Yukawa couplings and CKM mixing elements \cite{Antusch:2025fpm}
\begin{align}
 \label{input}
y_u = 2.87\times 10^{-6}, \, \,
y_d = 6.5\times 10^{-6}, \, \,
y_s = 1.29\times 10^{-4}, \, \, 
y_c = 1.45\times 10^{-3}, \, \, \nonumber \\
y_b = 6.06\times 10^{-3},\,\, 
y_t = 0.445, \,\, V_{us}=0.227, \,\, 
V_{cb}=4.74\times10^{-2},\,\, \\
V_{ub}=4.19\times 10^{-3}. \nonumber
\end{align}
to determine coefficient in front of $\epsilon$ parameter in the up and down quark mass matrices presented in Eq.~(\ref{matr1}).
In our fit, we choose $\epsilon =0.25$ and set the overall coefficient (see Eq.~(\ref{matr1})) of the down-quark mass matrix to $\epsilon^{3}\langle H\rangle = 174\, \epsilon^3 \approx 2.72 $.
\begin{eqnarray}
\label{matr2}
 M_U&= 174 &\left(
\begin{array}{ccc}
 0.19 \epsilon^{8} & 0.1\epsilon^{6} & 0.1 \epsilon^{4} \\
 0.1 \epsilon^{6} & 0.37 \epsilon^{4} & 0.1\epsilon^{2} \\
 0.1 \epsilon^{4} & 0.1\epsilon^{2} & 0.45 \\ 
\end{array}
\right),
~~~~ M_D= 2.72 \left(
\begin{array}{ccc}
 0.43 \epsilon^{5} & 0.49 \epsilon^{4} & 0.44\epsilon^{4} \\
 0.1 \epsilon^{3} & 0.13 \epsilon^{2} & 0.30 \epsilon^{2} \\
 0.1 \epsilon & 1 & 0.39 \\ 
\end{array}
\right).
\end{eqnarray}
We consider a scenario where all right-handed neutrinos have a universal mass, $M_M \approx 10^{12}$ GeV. The precise values for $M_{M}$ are obtained after performing a fit for neutrino masses and mixing. We adopt the type-I seesaw mechanism to generate light neutrino masses and mixing.
We used the values \cite{Antusch:2025fpm} for the charged-lepton Yukawa couplings, PMNS mixing matrix elements, and the solar $(\Delta m_{21}^{2}$) and atmospheric $(\Delta m_{31}^{2})$ mass splittings at $10^{12}$ GeV, which were generated via two-loop RGE running using PDG data
\begin{align}
 \label{input-2}
y_e = 2.80\times 10^{-6}, \, \,
y_{\mu} = 5.89\times 10^{-4}, \, \,
y_{\tau} = 1.00\times 10^{-2}, \, \, \nonumber \\
\theta_{21} = 33.4^o, \, \, \Delta m^2_{21} = 7.48\times 10^{-5} eV^2, \nonumber \\
\theta_{23} = 49.1^o, \, \, \Delta m^2_{31} = 2.53\times 10^{-3} eV^2 \\
\theta_{13} \approx 8.6^o. \nonumber
\end{align}
This allows us to determine the coefficients in front of the $\epsilon$ parameter in the charged-lepton and neutrino mass matrices presented in Eq.~(\ref{matr1}).

By performing a numerical fit, we obtain the charged-lepton and light-neutrino mass matrices presented below, which exhibit correct fit to the experimental data given above.
\begin{eqnarray}
\label{matr3}
M_e & = 2.72 &\left(
\begin{array}{ccc}
 0.18\epsilon^{5} & 0.9\epsilon^{3} & 0.38\epsilon \\
 0.45 \epsilon^{4} & 0.6 \epsilon^{2} & 0.5 \\
 0.35\epsilon^{4} & 0.4 & 0.64 \\ 
\end{array}
\right),
~~~~~ M^{lighr}_{\nu}= 0.05 \left(
\begin{array}{ccc}
 0.9\epsilon^{2} & 0.78\epsilon & 0.4\epsilon \\
 0.6\epsilon & 0.8 & 1 \\
 0.4\epsilon & 1 & 1 \\
\end{array}
\right) 
\end{eqnarray}
Here the overall coefficient of the charge 
lepton mass matrix to $\epsilon^{3}\langle H\rangle = 174\, \epsilon^3 \approx 2.72$. We obtain the following overall coefficient for the light neutrino mass matrix: $\frac{\langle H\rangle^2}{M_M} \epsilon^{2p} = \frac{\langle H\rangle^2}{2.36\times 10^{12}} \epsilon^{4}  \approx 0.05$.
The values $p=2$ and $M_{M}=2.36\times 10^{12}$ are selected in Eq.~(\ref{matr1}) to accurately reproduce the observed solar and atmospheric neutrino mass splitting.

The scalar potential allowed by the $SU(2)_L\times U(1)_Y\times U(1)_F$ symmetry is
\begin{equation}
\label{pot-1}
\begin{aligned}
V(H,\Phi,S) =- m_1^2 H^\dagger H + \lambda (H^\dagger H)^2-m_2^2\Phi^*\Phi + \lambda_1(\Phi^*\Phi)^2 -
m_3^2 S^*S + \lambda_2 ( S^*S)^2 \\
+ \lambda_3(\Phi^*\Phi)(S^*S) + \lambda_4 (H^\dagger H)(S^*S)
+\lambda_5(H^\dagger H)(\Phi^*\Phi)+ (m_{12}\Phi^*S^2+h.c.)
\end{aligned}
\end{equation}
All mass squares and quartic couplings in the potential are strictly real-valued. The parameter $m_{12}$ is complex.
Note that $U(1)_F$ charges for complex scalar fields $\Phi$ and $S$ are $q_{\Phi}=2$ and $q_S=1$ respectively.

We consider a scenario in which the complex scalar singlet field $\Phi$ first develops a vacuum expectation value (VEV), $\langle \Phi\rangle\equiv v_F/\sqrt{2}$, at a high scale. Consequently, the flavor symmetry $U(1)_F$ is spontaneously broken $U(1)_F \rightarrow Z_2$. As a result, the $U(1)_F$ gauge boson ($Z^{\prime}$) and the physical component of complex scalar $\Phi$ field become massive, $O(v_F)$. Typically for $v_F >10^{11}$ GeV, the contributions of the $Z^{\prime}$ and $\Phi$ fields to low-scale observables are negligible \cite{Blasi:2024vew}. 

In our scenario, the $\Phi$ field develops a VEV at a much higher scale compared to the VEVs $\langle S\rangle\equiv v_{Z_{2}}$ and $\langle H\rangle\equiv v_F$, ($v_F\gg v_{Z_{2}}, v_{\rm SM}$).
After the heavy field $\Phi$ develops a VEV, we can derive the effective potential $V(H, S)$ for the SM Higgs doublet $(H)$ and complex scalar $(S)$ by substituting $\Phi \rightarrow (v_{F}+\phi )/\sqrt{2})$ 
and integrating out the heavy $\Phi$ at an energy scale below $v_F$ we obtain:
\begin{equation}
\label{pot-2}
\begin{aligned}
V(H, S) = & - \mu^2H^\dagger H+\lambda (H^\dagger H)^2 - m^2 S^*S +  
+ \lambda_2 (S^* S)^2 \\ & + \lambda_3 (H^\dagger H)(S^*S) 
+ \frac{m_{12} v_F}{{2}}(S^2+(S^*)^2)
\end{aligned}
\end{equation}
We assume that all dimensionless quartic couplings $\lambda$, $\lambda_2$ and $\lambda_3$ are real. $m_{12}$ denotes a complex mass parameter. 
We adopt the following notation:
\begin{equation}
\label{pot-3}
\begin{aligned}
\mu^2 = M_1^2 -\frac{1}{2}\lambda_3 v_F^2 \\
m^2 = M_3^2-\frac{1}{2}\lambda_5 v_F^2
\end{aligned}
\end{equation}
In our model, below the $v_{F}$ scale, we have effectively an extension of the SM with a complex scalar field carrying a discrete $Z_{2}$ symmetry \cite{Barger:2007im, Profumo:2014opa,Godunov:2015nea}.

We assume that the SM Higgs doublet acquires a VEV of $(v=246\text{\ GeV})$ following spontaneous symmetry breaking. In our model, the VEV for the $S$ field can lie anywhere in the range $(v_{Z_{2}}<v_{F})$. 

Here we consider non-decoupling regime for $H$ and $S$ fields:

\begin{eqnarray}
\label{vev-5}
H=\frac{1}{\sqrt{2}}\left(
\begin{array}{c}
 0 \\
  v+h \\
\end{array}
\right), \hspace{3cm}  S=\frac{v_{Z_{2}}+s+ia}{\sqrt{2}}
\end{eqnarray}
The CP-odd scalar singlet mass is 
\begin{eqnarray}
\label{cp-odd}
m_a^2=-m_{12}v_{Z_{F}}.
\end{eqnarray}
The resulting mass-squared matrix for the CP-even components fluctuating about the VEVs is given by
\begin{eqnarray}
\label{higgs-scalar}
\left(
\begin{array}{cc}
 2\lambda v^2& \lambda_3v^2v^2_{Z_{2}} \\
  \lambda_3v^2v^2_{Z_{2}} & 2\lambda_2 v^2_{Z_{2}} \\
\end{array}
\right)
\end{eqnarray}
CP-even Higgs physical masses are 
\begin{eqnarray}
\label{higgs-scalar-2}
m_{1,2}^2 = \lambda v^2+\lambda_2v_{Z_{2}} - (\lambda v^2+\lambda_2v_{Z_{2}})\sqrt{1+\left(\frac{\lambda_3 v v_{Z_{2}}}{\lambda v^2-\lambda_2v_{Z_{2}}}\right)}
\end{eqnarray}
here we impose that $m_{1} = 125$ GeV. 
The mixing between the SM Higgs boson and single scalar field is 
\begin{eqnarray}
\label{mix-2}
\tan(\theta)=\frac{\lambda_3 v v_{Z_{2}}}{\lambda v^2-\lambda_2v^2_{Z_{2}}}.
\end{eqnarray}

Due to the gauged $U(1)_F$ symmetry, in the model, a specific assignment of $U(1)_F$ charges must be given to the chiral fermions such that all local gauge anomalies—including mixed $U(1)_F$ and gauge-gravitational contributions—vanish \cite{Adler:1969gk, Bell:1969ts}. The $U(1)_F$ charge assignment presented in Table~\ref{tcharge} leads us to the chiral anomalies in the theory.
To ensure anomaly cancellation, we must introduce additional chiral fermions. These are chosen to cancel all chiral anomalies involving the $U(1)_{F}$ symmetry
while remaining anomaly-free with respect to the SM gauge symmetry. 

Furthermore, these new fermions are expected to acquire mass once the $U(1)_{F}$ symmetry is broken. In general, ensuring the gauge invariance of a $U(1)_{F}$ symmetry is a non-trivial task. 
It was shown \cite{Dudas:1995yu, Chen:2008tc, Allanach:2018vjg, Bonnefoy:2019lsn, Tavartkiladze:2025oiq, Tavartkiladze:2022pzf, Babu:2026yqp} that it is possible to find a set of chiral fermions that render the $U(1)_F$ symmetry gauge invariant, while simultaneously ensuring that all additional fermion masses lie around the $U(1)_F$ symmetry breaking scale.
In this paper we will not present explicitly anomaly-free set of additional particle spectrum. Instead, we assume that all of them acquire mass upon the spontaneous breaking of the $U(1)_{F}$ symmetry. For the $U(1)_{F}$ symmetry breaking scale $v_F$ above $10^{11}$ GeV, the new particles associated with this symmetry do not contribute to electroweak physics tests \cite{Blasi:2024vew,ParticleDataGroup:2024cfk}.

Note that, as shown in \cite{Bonnefoy:2019lsn}, the introduction of additional chiral fermions to gauge the $U(1)_F$ symmetry—with all such particles residing near the symmetry-breaking scale--causes the gauge couplings to become non-perturbative shortly above that scale.

\section{Dark Matter}

In this subsection, we briefly discuss the dark matter candidate that naturally arises in our model. The potential in Eq.~\eqref{pot-2} has a softly broken global $U(1)$ symmetry (the last term). After the SM Higgs field $H$ and the complex scalar $S$ develop VEVs, the CP-odd component of the $S$ field, denoted as $a$, can be identified as a pseudo-Nambu-Goldstone boson (pNGB) (see Eq.~\eqref{cp-odd}). Since the SM fields are even under this $Z_2$ symmetry and the pseudo-scalar $a$ field cannot mix with real scalars, it cannot decay into any SM final states. This makes the pseudo-scalar $a$ field an excellent candidate for dark matter (DM) \cite{Barger:2008jx, Gonderinger:2012rd, Gross:2017dan}.

 In our model, the physical mass of the dark matter candidate (the CP-odd pNGB, $a$) is given by by the last term in the Eq.~\eqref{pot-2}:
\begin{eqnarray}
\label{cp-odd1}
m^2_{DM}=m_a^2=-m_{12}v_{Z_{F}}.
\end{eqnarray}
By choosing appropriate values for the $m_{12}$ mass parameter in Eq.~\eqref{cp-odd1}, we can treat the dark matter mass as a free, adjustable parameter spanning a wide range from a few GeV to tens of TeV. One possibility is to consider the DM mass $m_{\rm DM} \approx 62.5 ~\mathrm{GeV}$. In this case, dark-matter pairs $(aa)$ annihilate extremely efficiently into Standard Model particles by utilizing an s-channel resonance through the Higgs boson. Because the annihilation rate is enhanced by this resonance, one only needs a tiny Higgs-portal coupling ($\lambda_3(H^\dagger H)(S^*S)$; see Eq.~\eqref{pot-2}). 

A second possibility for the dark matter candidate in our scenario is that $m_a$ is heavier than the Higgs boson but below the multi-TeV scale. In this regime, annihilation channels like $aa \to W^+W^-$, $aa \to ZZ$, and $aa \to hh$ open up. The relic density is perfectly satisfied with 
$\lambda_3 \sim 0.1$. Its direct detection scattering amplitude is suppressed by the momentum transfer, keeping it well hidden from underground detectors.

A third possibility is $1 \mathrm{TeV} < m_a < 100 \mathrm{TeV}$. To keep dark matter annihilating efficiently enough so that it doesn't overclose the universe, one either needs a larger coupling $(\lambda_3)$ or a heavy CP-even scalar partner $(s)$ that generates an s-channel resonance, $(m_a\approx m_s/2)$. The upper bound $m_a < 100 \mathrm{TeV}$ or so can be derived from the unitarity and perturbativity conditions of the $\lambda _{3}$ couplings \cite{Griest:1989wd}.

The direct detection cross section is vanishingly small in the limit of zero momentum transfer due to its pseudo-Goldstone nature, which can alleviate the direct detection bounds \cite{PandaX-4T:2021bab,LZ:2022ufs,XENON:2023sxq} on WIMP-like dark matter. We do not discuss the detailed phenomenology of such CP-odd state DM as it is well studied in the literature \cite{Barger:2008jx, Gonderinger:2012rd, Gross:2017dan}, including in the context of the gauged $U(1)_{B-L}$ or general $U(1)_{X}$, see Refs.~\cite{Abe:2020iph,Okada:2020zxo} and possible embeddings in SO(10) \cite{Okada:2021qmi,Maji:2023fba}. Since direct-detection experimental cross sections are relevant at zero momentum transfer, the WIMP-nucleon scattering cross section becomes vanishingly small at tree level~\cite{Gross:2017dan}, evading most of the direct-detection constraints~\cite{Arcadi:2016qoz}.

\subsection*{Laboratory Search for CP-even state:} 

Constraints on the complex scalar $S$ are critically dependent on the $\lambda _{3}$ coupling and the $v_{Z_{2}}$ scale.
For the CP-even state, $h_2$, if the couplings are weak, there is an interesting way to search for these particles, as they may show signatures in long-lived particle (LLP) searches. In particular, long-lived particles with masses $m_2 \lesssim 5 \mathrm{GeV}$ may be detectable with upcoming experiments such as FASER and FASER-II \cite{Feng:2017vli,FASER:2018eoc,FASER:2018bac,FASER:2019aik}, DUNE \cite{DUNE:2015lol,Berryman:2019dme}, DarkQuest-Phase 2 \cite{Batell:2020vqn}, MATHUSLA \cite{Curtin:2018mvb}, PS191 \cite{Bernardi:1985ny,Gorbunov:2021ccu} or SHIP \cite{SHiP:2015vad} with typical $\sin(\theta)$ (determined by Eq.~\eqref{mix-2}) ranging between $\mathcal{O}(10^{-6})$--$\mathcal{O}(10^{-3})$ in these upcoming experimental probes (see Ref.~\cite{FASER:2019aik} for details) for $m_a$ around $100\,\mathrm{MeV}$--$10\,\mathrm{GeV}$.

\section{Topological Defects: Walls bounded by Strings}
We examine a scenario where the spontaneous breaking of $U(1)_F$ symmetry occurs as a second-order phase transition (see Eq.~(\ref{pot-1})). In order to avoid the sizable generation of a cubic term via radiative corrections, we assume all quartic couplings in Eq.~(\ref{pot-1}) satisfy the condition $\lambda_i \gg g_F$, where $i$ labels the quartic terms and $g_F$ is the gauge coupling corresponding to the $U(1)_F$ symmetry. Consequently, we focus exclusively on gravitational wave generation via the topological defects resulting from the spontaneous gauge symmetry breaking. 

Two realizations of $U(1)_F$ gauge symmetry breaking are possible. The first involves the direct breaking of $U(1)_F$ to the identity ($U(1)_F \rightarrow I$), in which case the first homotopy group of the vacuum manifold is non-trivial, $\pi_1(U(1)/1) \neq I$. Here $I$ stands for trivial homotopy group.  This symmetry-breaking pattern leads to the creation of a cosmic string network \cite{Kibble:1976sj} that emits gravitational waves \cite{Vachaspati:1984gt}. The second possibility is the symmetry-breaking chain $U(1)_F \rightarrow Z_N\rightarrow I$. 
The energy density of the network is quickly
dominated by horizon-size DWs.

For simplicity, we investigate the $U(1)_F \rightarrow Z_2\rightarrow I$ symmetry-breaking chain in this paper.
The $U(1)_F$ flavor symmetry is broken to a $Z_2$ subgroup by introducing a complex scalar field $\Phi$ with a flavor charge $q_{\Phi}=2$. Since the first homotopy group of the vacuum manifold $U(1)/Z_2$ is non-trivial, with $\pi_1(U(1)/Z_2) \simeq Z$, this produces cosmic strings.

 Once the $S$ field develops a vacuum expectation value (VEV), the remaining $Z_2$ symmetry is also broken. For the second stage of symmetry breaking, $Z_2 \rightarrow 1$, the zeroth homotopy group of the vacuum manifold $(Z_2/1)$ is non-trivial because it is disconnected, with $\pi_0(Z_2/1) \cong Z_2$.
 As discussed earlier, we find the flavor model naturally accommodates a hybrid topological defect containing domain walls bounded by local cosmic strings or simply, walls-bounded-by-strings. 
 As we will show in a later section, this hybrid defect leads to characteristic spectral features in the GW signal, allowing us to probe the $U(1)_F$ breaking scale up to $10^{15}$ GeV and, in some regions, complement laboratory flavor observables.

Assuming inflation occurred before string formation, a significant number of strings would be within the horizon at the time of wall formation. In this case, the space between strings is filled by domain walls, realizing the "walls bounded by strings" scenario \cite{Kibble:1982dd,Everett:1982nm,Lazarides:2023iim} at the $Z_2$
symmetry-breaking scale. These domain walls are unstable, as the full theory satisfies $\pi_0(U(1)/1) = I$.

The scalar field $\Phi$, which breaks $U(1)_F$ symmetry, couples to additional fields necessary for chiral anomaly cancellation. Since these fields reside at the $U(1)_F$
symmetry breaking scale, the resulting cosmic strings can carry both charged and neutral currents, characterizing them as superconducting strings \cite{Witten:1984eb}. Such superconducting strings can produce the GW spectrum compared to Nambu-Goto strings due to the presence of the fermion zero mode; see Refs.~\cite{ Sousa:2020sxs, Buchmuller:2021mbb,Rybak:2022sbo, Afzal:2023kqs}. Our model considers a superconducting domain wall bounded by superconducting strings. Superconducting domain walls also lead to unique two-peaked GW signatures as shown in Ref.~\cite{Ghoshal:2025gci}. As the dynamics crucially depend upon the $U(1)_F$ gauge coupling (string charge/current density), we assume the limit $g_F\ll 1$ and neglect current-related effects in the present gravitational wave spectrum calculation. A more comprehensive analysis will be provided in future work.

 To summarize, for the breaking chain 
 \begin{align}
  U(1) \stackrel{\langle \Phi \rangle}{\xrightarrow{\hspace{0.8cm}}} Z_2 \stackrel{\langle S \rangle}{\xrightarrow{\hspace{0.8cm}}} I. \label{UtoZ2}
\end{align}
 successive stages of symmetry breaking produce cosmic strings and then domain walls. Since the cosmic string network already exists when the domain walls appear, walls get attached to the strings. Note that this hybrid network, walls bounded by strings, features a signal with $f^{3}$ slope in the infrared frequencies, quite distinguishable from the pure string case \cite{Dunsky:2021tih, Maji:2023fba, Lazarides:2023ksx}.
In order to pinpoint the detectability of the GW spectrum arising from the hybrid defects, in this paper we identify a strategy to formulate a "turning-point threshold frequency" and its relative uncertainty that various GW detectors will be able to detect and compare this with the pure cosmic-string case.



\section{Gravitational Wave Gastronomy: Walls bounded by Strings}
A detailed description of the evolution of this hybrid topological defect can be found in Refs.~\cite{Dunsky:2021tih,Fu:2024rsm}. Here we will briefly review the relevant physics and the observational signatures of the defect. In order to consider a wall bounded by a string of radius of curvature, say $R$, the force per unit length on the string boundary $\sim \mu/R$ starts to dominate over the wall tension $\sigma$ for $R<R_c=\mu/\sigma$. One useful parameter that controls the evolution of this hybrid defect is defined as the critical radius $R_c \equiv \mu / \sigma$, where $\mu \sim v_{\rm F}^2$ is the cosmic string tension, $v_{\rm F}$ being the $U(1)_F$ breaking scale, and $\sigma \sim v_{\rm{Z_2}}^3$ is the surface energy density of the domain wall, $v_{\rm{Z_2}}$ being the $Z_2$ breaking scale. Due to the subsequent symmetry-breaking chain, $U(1)_F$ and then $Z_2$, non-relativistic cosmic string loops lose their energy by emitting gravitational waves until the time of DW formation $t_{\rm DW} \approx M_{\rm Pl} \sqrt{(8\pi^3 g_\star/ 90)} / v_{\rm{Z_2}}^2$. Here the loops become ultra-relativistic as long as they have the size $R_c$ and $t_{\rm DW} < R_c$ guarantees that they will not be the dominant component of the energy budget of the universe. As the collapse of the hybrid-string wall system now happens earlier than the pure string loops, it will exhibit a feature of departure from the standard cosmic string spectrum in the GW frequency spectrum with the collapse time being $t_\star \sim t_{\rm DW}$. Typically, the collapse time can be estimated as $t_\star \equiv \max(R_c, t_{\rm DW})$ \cite{Martin:1996ea}.

For a loop of radius $R$ with circular length of size $l = 2\pi R$, the rate of energy loss is given by
\begin{align}
  \frac{dE}{dt} = -\Gamma(l) G\mu^2,
\end{align}
where $G$ is Newton's constant. The function $\Gamma(l) \approx \Gamma_s = 50$ when $l \ll 2\pi R_c$ in the pure string limit, and $\Gamma(l) = 3.7(l/(2\pi R_c))^2$ for $l \gg 2\pi R_c$ in the pure wall limit. The behavior around $l \sim 2\pi R_c$ can be approximated using a smooth interpolation function, as shown in Ref.~\cite{Dunsky:2021tih}.

The relic Gravitational Wave energy density in the present universe with frequency $f$ in the present time,
\begin{align}
\Omega_{\rm GW} = \frac{1}{\rho_c}\frac{d\rho_{\rm GW}}{d\log f}
\end{align}
 where $\rho_c = 3H_0^2/(8\pi G)$ is the critical energy density of the present universe. The stochastic gravitational wave background from domain walls bounded by strings is estimated in Ref.~\cite{Dunsky:2021tih} as a sum of the contributions from normal modes given by:
\begin{align}
  \label{eq:GWs1}
\Omega_{\rm GW}(f)=\sum_n\Omega^{(n)}_{\rm GW} ,
\end{align}
where
\begin{align} 
  \label{eq:GWs2}
  \Omega^{(n)}_{\rm GW}
  = \frac{1}{\rho_c} \int_{t_F}^{t_0} d\tilde{t} \left(\frac{a(\tilde{t})}{a(t_0)}\right)^5\frac{\mathcal{F} C_{\rm eff}(t_i)}{\alpha t_i^4} \left(\frac{a(t_i)}{a(\tilde{t})}\right)^3
   \frac{\left(1 + \frac{\xi n}{2\pi R_c f}\frac{a(\tilde{t})}{a(t_0)} \right)}{\Gamma G \mu + \alpha(1 + \frac{\alpha t_i}{2\pi R_c})}\frac{\Gamma n^{-q}}{\zeta(q)} G\mu^2 \frac{\xi n}{f}\theta(t_* - t_i) ,
\end{align}
and a loop of initial length $l_i=\alpha t_i$ decays with its length ($l$) at any subsequent time $t$ given by
\begin{align}
  \label{eq:lengthLoss}
  G\mu(t - t_i) = \int_l^{\alpha t_i} dl' \frac{1 + \frac{l'}{2 \pi R_c}}{\Gamma(l')}.
\end{align}
 In our case, $t_* = R_c\gg t_{\rm dw}$, and therefore we have $\xi = 2$ and $\Gamma\simeq 50$ corresponding to the pure string limit \cite{Vachaspati:1984gt,Vilenkin:2000jqa}. We have taken $\mathcal{F} \simeq 0.1$, $\alpha\simeq 0.1$, $C_{\rm eff} = 5.7$ for the radiation-dominated universe \cite{Vanchurin:2005pa,Ringeval:2005kr,Olum:2006ix,Blanco-Pillado:2013qja,Blanco-Pillado:2017oxo,Cui:2018rwi}, and $q=4/3$ because of cusp domination in the gravitational wave spectrum \cite{Olmez:2010bi, Auclair:2019wcv, Cui:2019kkd, LIGOScientific:2021nrg}. We have included the sum of normal modes up to $n=10^4$.


Because of the energy loss, the cosmic string loops which forms at time $t_k$ with an initial size $l_k = \alpha t_k$ slowly decrease in size with time. Here $\alpha \approx 0.1$ is the ratio between loop formation length and horizon size, and its value is determined from simulations \cite{Blanco-Pillado:2017oxo, Blanco-Pillado:2013qja}. For the domain walls bounded by strings, the size of a string loop emitting gravitational waves at time $\tilde{t}$ is $\tilde{l} = 2\pi \tilde{R}$, and we have
\begin{align}
  G\mu (\tilde{t}-t_k) = \int_{\tilde{l}}^{\alpha t_k} d l' \frac{1+\frac{l'}{2\pi R_c}}{\Gamma(l')}. \label{tksolve}
\end{align}
Here the integration variable $l'$ can be interpreted as the instantaneous length of the loop emitting gravitational waves at time $t'$, where $t_k < t' < \tilde{t}$. We further assume that at time $\tilde{t}$, the newly created loops are larger than the pre-existing loops emitting gravitational waves at time $\tilde{t}$
\begin{align}
  \alpha \tilde{t} > \tilde{l}.
\end{align}

The energy density of the stochastic gravitational wave spectrum at time $t$ is given by
\begin{align}
  \frac{d\rho_{\rm GW}(t)}{df} &= \int_{t_{\rm sc}}^{t} d\tilde{t} \left(\frac{a(\tilde{t})}{a(t)}\right)^4 \int dl \frac{dn(l, \tilde{t})}{dl} \frac{dP(l, \tilde{t})}{d\tilde{f}} \frac{d\tilde{f}}{df}. \label{drhodf}
\end{align}
Here $\tilde{t}$ is the emission time of the gravitational wave, and $t_{\rm sc} \sim 10^{-22}$ s is the time when the network reaches a scaling regime where the energy density of the hybrid network becomes constant in the expanding Universe. 
The first term of the rightmost integrand of Eq.~\eqref{drhodf} can be decomposed as 
\begin{align}
  \frac{dn(l, \tilde{t})}{dl} = \frac{dn}{dt_k} \frac{dt_k}{dl}. \label{dndl}
\end{align}
The second term in Eq.~\eqref{dndl} is determined by differentiating Eq.~\eqref{tksolve} with respect to $t_k$,
\begin{align}
  \frac{dt_k}{dl} = \frac{1+\frac{l}{2\pi R_c}}{\Gamma(l) G\mu} \left[ 1 + \frac{\alpha \left( 1 + \frac{\alpha t_k}{2\pi R_c}\right)}{\Gamma(\alpha t_k)G \mu} \right]^{-1},
\end{align}
while the first term represents the loop number density production rate calculated from the one-scale model and calibrated from simulations \cite{Cui:2018rwi, Gouttenoire:2019kij, Sousa:2013aaa}
\begin{align}
  \frac{dn}{dt_k} &= \left[ \frac{\mathcal{F} C_{\rm eff}(t_k)}{\alpha t_k^4} \left(\frac{a(t_k)}{a(\tilde{t})}\right)^3 \right] \theta{(t_\star - t_k)}.
\end{align}
We have assumed that at every Hubble time, roughly one loop of size $\alpha t_k$ breaks off from the infinite wall-string network, and then it redshifts as $a^{-3}$, $a(t)$ being the scale factor. $\mathcal{F} \approx 0.1$ is the fraction of energy that is transferred by the string network into loops \cite{Blanco-Pillado:2013qja}. $C_{\rm eff}$ is the loop formation efficiency which equals $5.7$ during the radiation-dominated era, and $0.5$ during the matter-dominated era \cite{Cui:2017ufi, Blasi:2020wpy}. The Heaviside $\theta$ function $\theta(t_\star - t_k)$ ensures that we only consider the contribution from loops which were created before the network collapses at $t_\star$.

The second term in the rightmost integrand of Eq.~\eqref{drhodf} can be expressed as
\begin{align}
  \frac{dP(l, \tilde{t})}{d\tilde{f}} &= \Gamma(l) G\mu^2l\ g(\tilde{f}l).
\end{align}
It is useful to decompose the gravitational wave into Fourier modes $\tilde{f}_n = \xi n/\tilde{l}$, where $\xi \equiv l/T$ varies between $2$ in the pure string limit ($\tilde{l} \ll 2\pi R_c$) and $\pi$ in the pure wall limit $\tilde{l} \gg 2\pi R_c$ \cite{Dunsky:2021tih}.. The normalized power spectrum for a discrete spectrum is then given by
\begin{align}
  g(x) &= \sum_n \mathcal{P}_n\ \delta(x - \xi n). \label{gx}
\end{align}
Here $\mathcal{P}_n = n^{-q}/\zeta(q)$ is the fractional power radiated by the $n$th mode of an oscillating string loop with cusp, and $\zeta(x) = \sum_{m=1}^{\infty} m^{-x}$ is the Riemann zeta function. The power spectral index is found to be $q = 4/3$ for string loops containing cusps \cite{Auclair:2019wcv, Vachaspati:1984gt}. 

Finally, accounting for redshifting, the emission frequency $\tilde{f}$ at time $\tilde{t}$ is related to the observed frequency $f$ at time $t$ by
\begin{align}
  f = \tilde{f}\ \frac{a(\tilde{t})}{a(t)}, \label{fredshift}
\end{align}
where the scale factor $a(t)$ is defined as $a(t_0) \equiv 1$ for present time $t_0$. From Eq.~\eqref{fredshift} the third term of the rightmost integrand in Eq.~\eqref{drhodf} can be easily obtained.

Putting everything together, the stochastic gravitational wave spectrum observed today is given by summing over all Fourier modes
\begin{align}
  \Omega_{\rm GW} (f) &\equiv \frac{f}{\rho_c} \frac{d\rho_{\rm GW}}{df} \nonumber \\
  &= \sum_n \frac{G\mu^2}{\rho_c} \int_{t_{\rm sc}}^{t_0} d\tilde{t} \left( \frac{a(\tilde{t})}{a(t_0)}\right)^5 \left[ \frac{\mathcal{F} C_{\rm eff}(t_k)}{\alpha t_k^4} \left(\frac{a(t_k)}{a(\tilde{t})}\right)^3 \right] \mathcal{P}_n \frac{\xi n}{f} \left[ 1+ \frac{1}{2\pi R_c } \frac{\xi n}{f}\frac{a(\tilde{t})}{a(t_0)} \right] \nonumber \\
  &\times \frac{\Gamma(\alpha t_k) \theta{(t_\star - t_k)}}{\Gamma(\alpha t_k) G\mu + \alpha \left(1+\frac{\alpha t_k}{2\pi R_c}\right)}. \label{GWformula}
\end{align}

Here $\rho_c = 3H_0^2/(8\pi G)$ is the critical energy density today. The integral with respect to $l$ in Eq.~\eqref{drhodf} is performed utilizing the $\delta$ function in Eq.~\eqref{gx} to replace $l = \frac{\xi n}{f} \frac{a(\tilde{t})}{a(t_0)}$ for each $n$-mode.

In Eq.~\eqref{GWformula}, we are integrating over the gravitational wave modes emitted by the hybrid network from an early time when the network goes to a scaling regime, $t_{\rm sc}$, to today $t_0$. Here the integration variable $\tilde{t}$ represents the time when a loop is radiating gravitational wave, and $t_k$ is the time when this particular loop was created, which can be obtained by solving Eq.~\eqref{tksolve}.  

If no domain walls were created (when the $U(1)_F$ symmetry is broken to nothing), we could take the $R_c \rightarrow \infty$ limit, and Eq.~\eqref{eq:GWs2} would yield the gravitational wave spectrum generated from the decay of a pure string network. In this case, the Heaviside theta function would be replaced by $\theta(t_k - t_{\rm sc})$ to imply that we only consider the loops created after $t_{\rm sc}$ when the network reaches a scaling regime, and loop creation does not stop at any characteristic time similar to $t_{\star}$. This distinction has important consequences for the infrared tail of the GW spectrum, as will be discussed later.

\section{Characteristics of the SGWB}

We calculate the gravitational wave spectrum from the string-wall network (solid lines), calculated using Eq.~(\ref{GWformula}) up to $k=10^4$ modes. We also include the pure string scenario ($R_c \to \infty$) in our plots, for comparison, representing the case where the $U(1)_F$ symmetry is completely broken at the high scale. To contextualize our results, we compare them against existing limits and future detection benchmarks. Current constraints include bounds from LVK and recent data from the NANOGrav \cite{NANOGrav:2023gor, NANOGrav:2023hvm} and EPTA \cite{EPTA:2023sfo, EPTA:2023fyk} pulsar timing arrays. Future experimental reaches are represented by projected sensitivities from (SKA \cite{Janssen:2014dka}, $\mu$-Ares \cite{Sesana:2019vho}, LISA \cite{LISA:2017pwj}, DECIGO \cite{Kudoh:2005as, Kawamura:2020pcg}, BBO \cite{Harry:2006fi}, AEDGE \cite{AEDGE:2019nxb}, AION \cite{Badurina:2019hst}, CE \cite{LIGOScientific:2016wof}, ET \cite{Hild:2008ng}, and next-generation LVK \cite{LIGOScientific:2022sts, KAGRA:2021kbb, Jiang:2022uxp} upgrades.

\begin{figure}[H]
  \centering
  \includegraphics[width=0.84\linewidth]{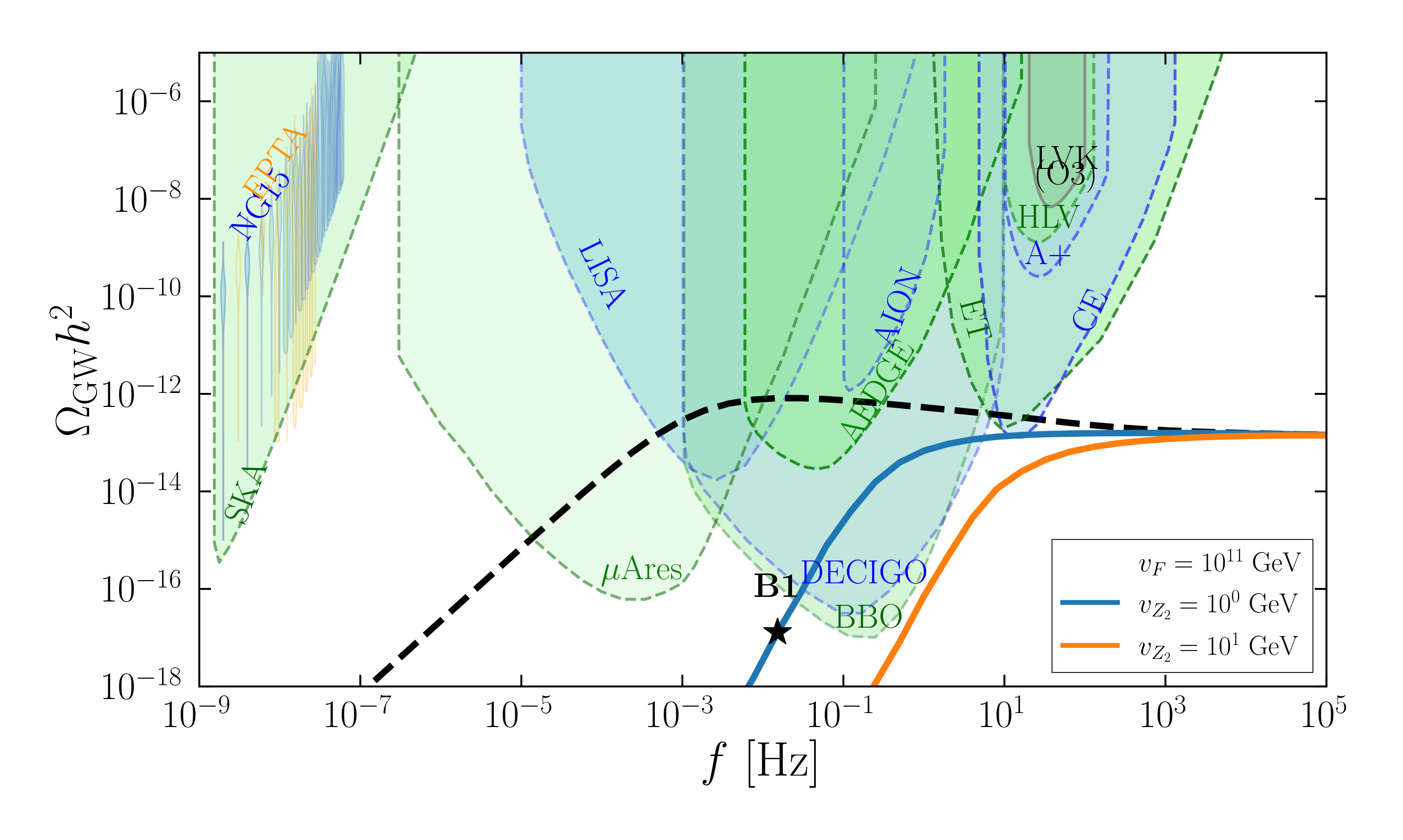}
  \includegraphics[width=0.84\linewidth]{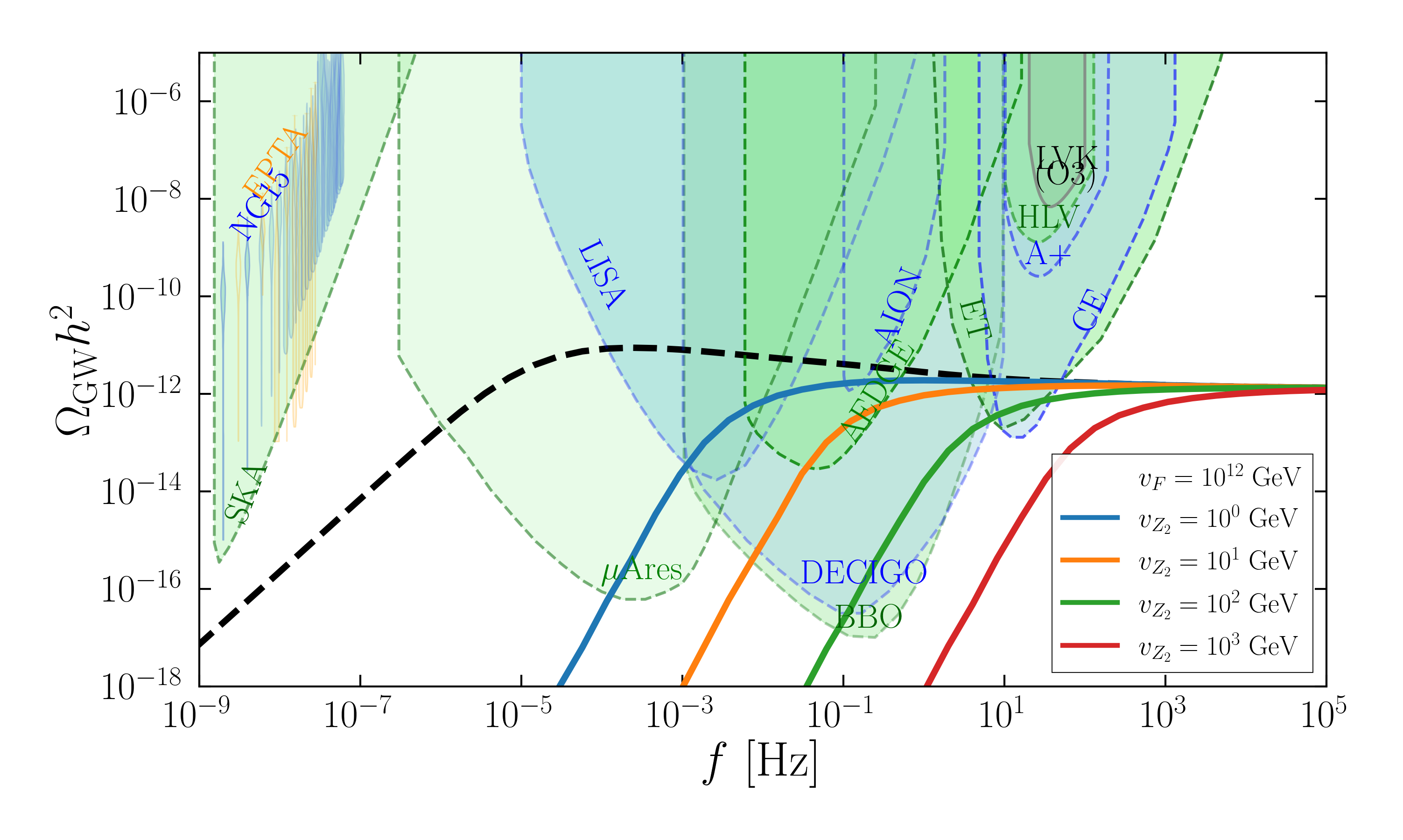}
  \caption{\it GW spectra from domain walls bounded by cosmic strings (solid colored curves) for two representative choices of the $U(1)_F$ breaking scale: $v_F = 10^{11}\,\mathrm{GeV}$ (top panel) and $v_F = 10^{12}\,\mathrm{GeV}$ (bottom panel), with various intermediate $Z_2$ breaking scales $v_{Z_2}$ as indicated in the legends. The black dashed curve in each panel shows the corresponding pure cosmic-string spectrum. Projected sensitivities of various upcoming interferometers are shown by green dashed lines. The solid curve in the upper-right corner shows the upper bound from the third observing run of the LIGO--Virgo--KAGRA Collaboration. The star labeled B1 denotes the benchmark spectrum also shown in the parameter scan in Fig.~\ref{fig:tp_summary}.}
  \label{fig:GW-spectrum}
\end{figure}

\begin{figure}[H]
  \centering
  \includegraphics[width=0.84\linewidth]{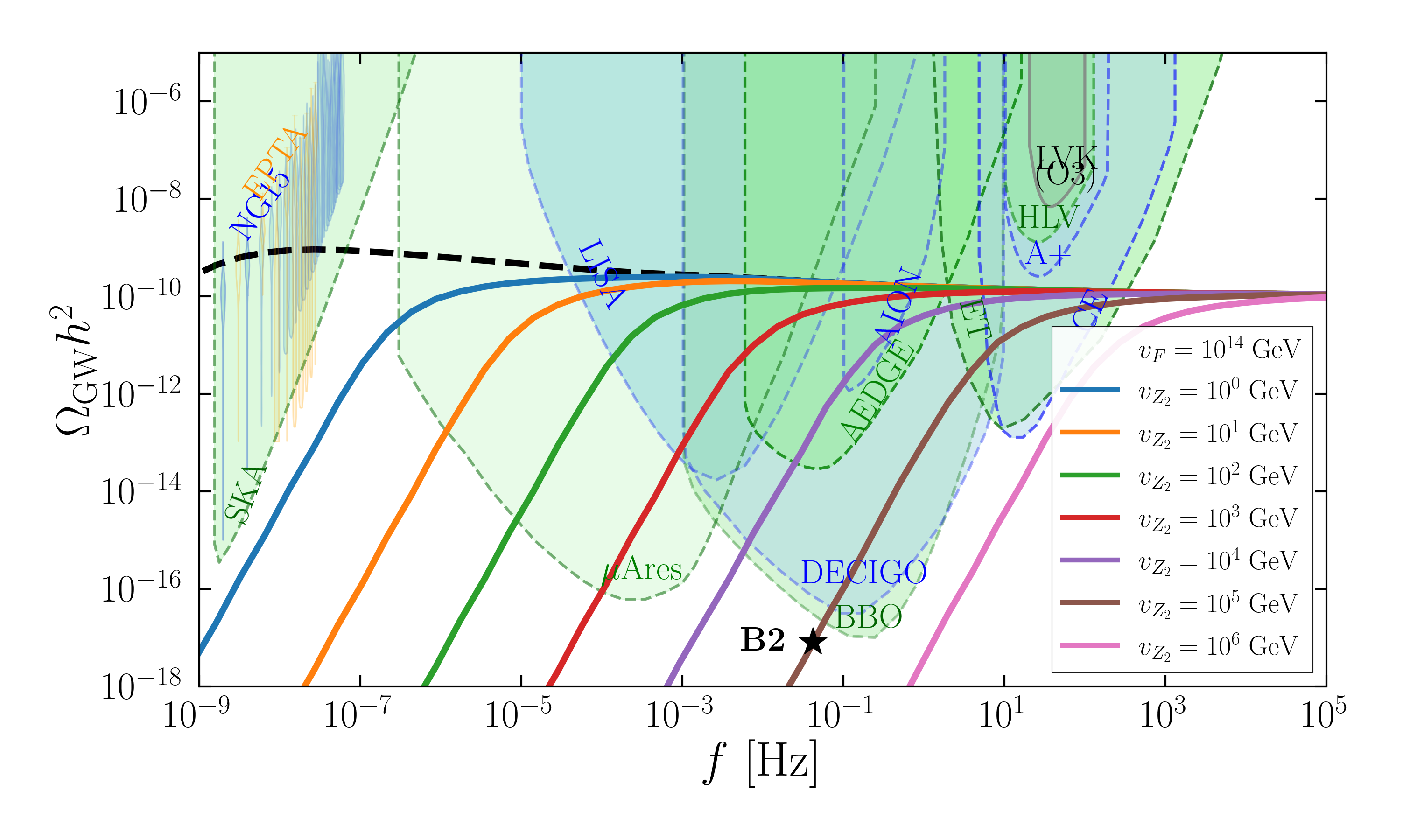}
  \includegraphics[width=0.84\linewidth]{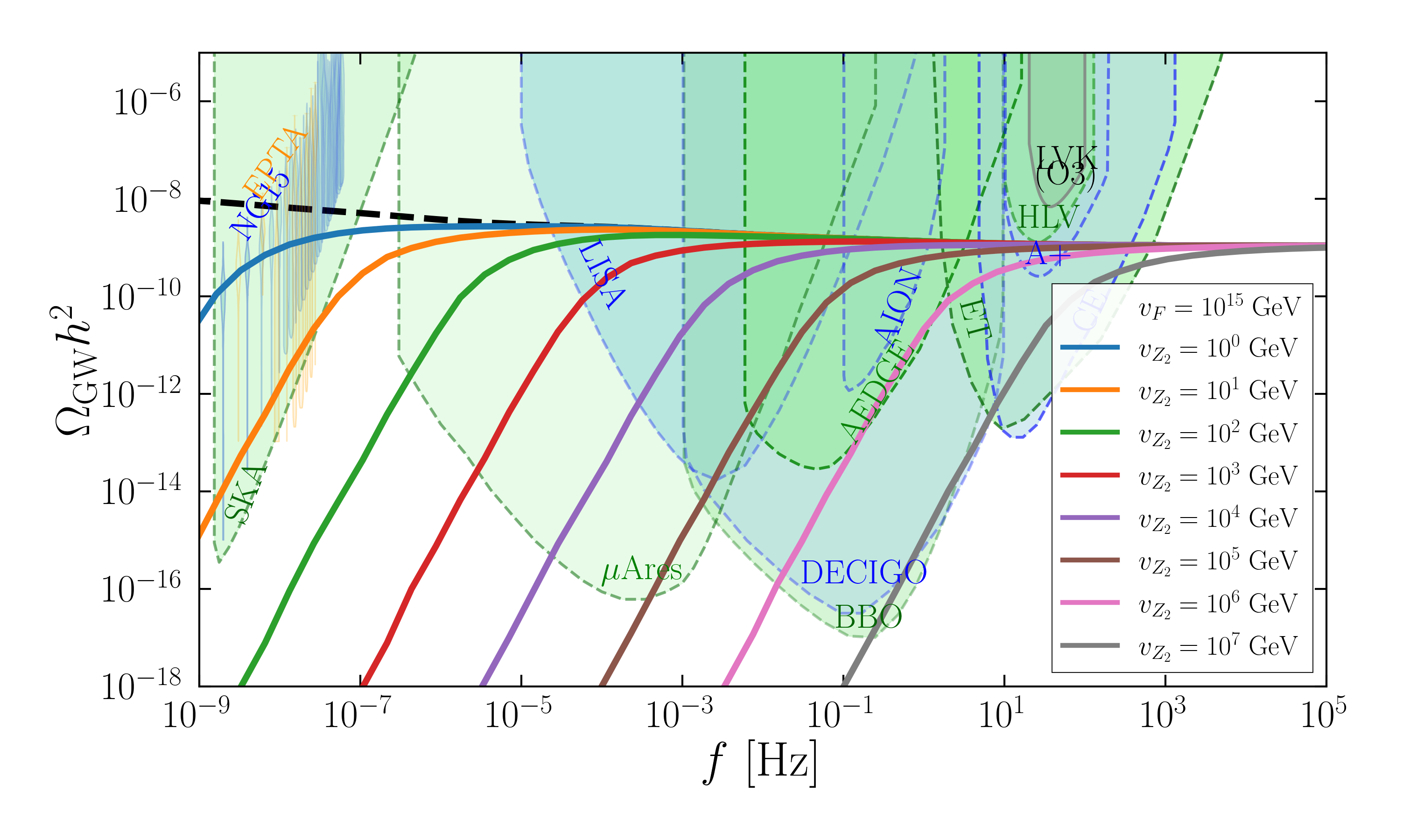}
  \caption{\it GW spectra from domain walls bounded by cosmic strings (solid colored curves) for two representative choices of the $U(1)_F$ breaking scale: $v_F = 10^{14}\,\mathrm{GeV}$ (top panel) and $v_F = 10^{15}\,\mathrm{GeV}$ (bottom panel), with various intermediate $Z_2$ breaking scales $v_{Z_2}$ as indicated in the legends. The black dashed curve in each panel shows the corresponding pure cosmic-string spectrum. Projected sensitivities of various upcoming interferometers are shown by green dashed lines. The solid curve in the upper-right corner shows the upper bound from the third observing run of the LIGO--Virgo--KAGRA Collaboration. The star labeled B2 denotes the benchmark spectrum also shown in the parameter scan in Fig.~\ref{fig:tp_summary}.}
  \label{fig:GW-spectrum-1}
\end{figure}

Let us discuss some characteristic features of the GW spectra shown in Figs. \ref{fig:GW-spectrum} and \ref{fig:GW-spectrum-1}. In contrast to the flat spectrum of the pure cosmic string case (dotted line), the signals from the hybrid network (solid line) feature an infrared (IR) tail. This tail clearly depicts a departure from the standard cosmic string plateau at lower frequencies. As discussed previously, this behavior arises because the strings decay via the domain walls, forcing the bounded strings to become ultra-relativistic. The frequency where this IR departure begins is determined by the collapse time $t_{\mathrm{D}W}$, which scales with the critical radius $R_c \sim v_{\rm F}^3 / v_{Z_2}^2$ (since $R_c > t_{\rm DW}$ for a fixed $v_{\mathrm{F}}$). Consequently, for a fixed $v_{Z_{2}}$, a larger critical radius $R_{c}$ corresponds to a higher $U(1)_F$ breaking scale. This implies that the hybrid defects collapse later in cosmic time, allowing the flat portion of the signal to extend to lower frequencies (see Figs. \ref{fig:GW-spectrum} and \ref{fig:GW-spectrum-1})
  
Unlike the pure cosmic-string spectrum, the hybrid-defect signal exhibits a steeper infrared slope characterized by an $f^{3}$ power law. This scaling is expected from causality arguments for sources that undergo rapid or instantaneous decay \cite{Caprini:2009fx, Cai:2019cdl, Brzeminski:2022haa}, and its physical origin lies in the dynamics of the string-bounded walls discussed above. As in the pure cosmic-string scenario, the early-time dynamics of the hybrid network, corresponding to higher observational frequencies, are dominated by non-relativistic string loops. This period captures the epoch after $U(1)_F$ breaking but before the discrete $Z_2$ phase transition. Once the $Z_2$ symmetry breaks, domain walls emerge and become bounded by the pre-existing string loops. The characteristic formation time of these walls, $t_{\rm DW}$, is governed by the $Z_2$ breaking scale $v_{Z_2}$. Provided that $t_{\rm DW}<R_c$, the newly formed walls do not immediately alter the ongoing string-dominated dynamics. The critical radius $R_c\sim v_F^3/v_{Z_2}^2$ is instead controlled by the hierarchy between the two symmetry-breaking scales. At the characteristic epoch $t_\star\equiv \max(R_c,t_{\rm DW})$, which reduces to $t_\star\sim R_c$ for $v_F\gtrsim 10^{10}$ GeV, two effects determine the infrared behavior of the resulting GW spectrum. First, the formation of new loops stops, leaving only the pre-existing loop population to decay. Second, the remaining loops are driven into an ultra-relativistic regime, which accelerates their fragmentation and leads to near-instantaneous decay. The overall dynamics therefore resemble a rapidly dissipating causal source, yielding the $f^3$ spectral slope. This pronounced infrared signature provides a clear observational handle for distinguishing the hybrid model from the pure-string baseline within planned $\mu$Hz-to-Hz interferometers.

\begin{figure}[H]
  \centering
  \includegraphics[width=0.78\linewidth]{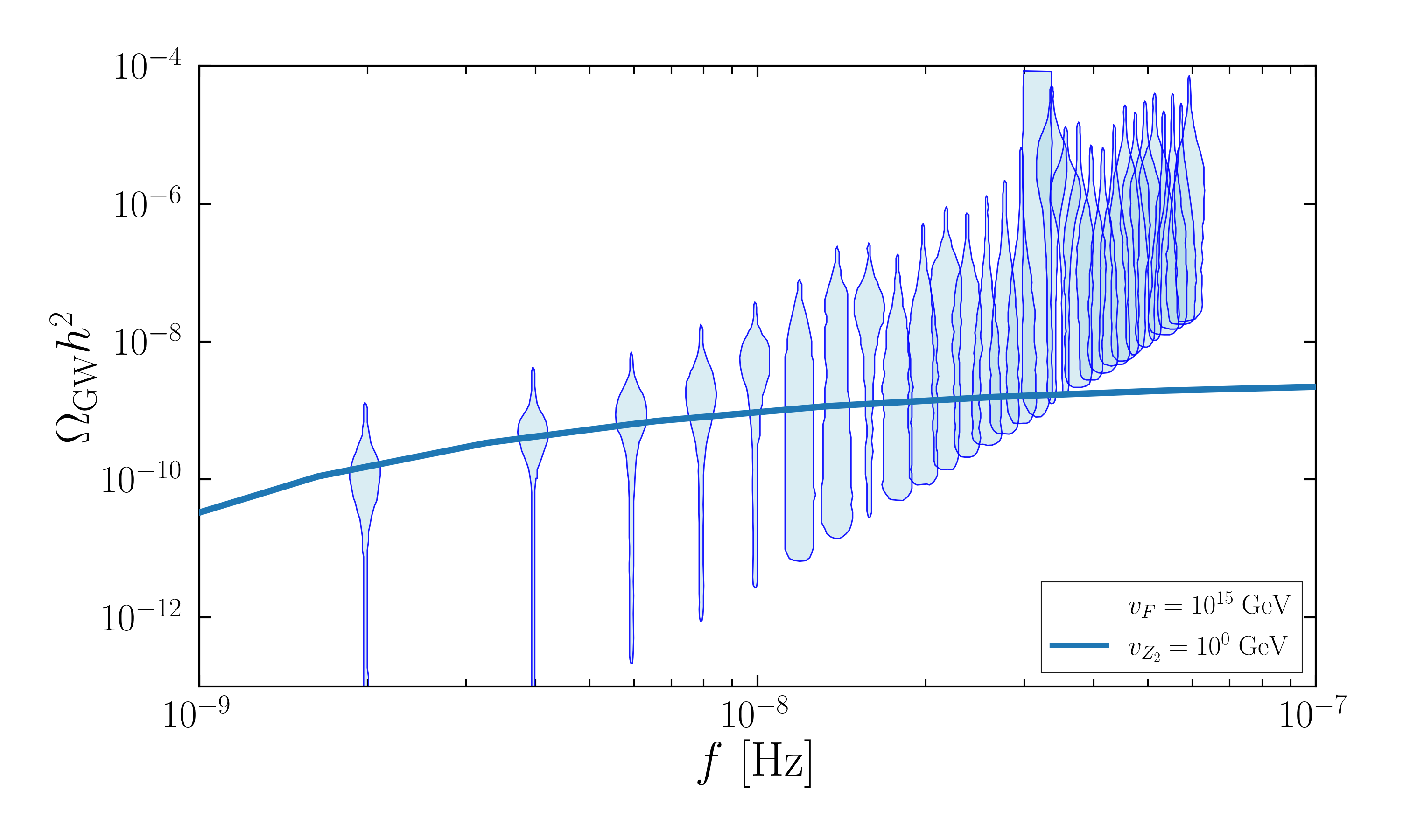}
  \caption{\it GW spectrum from domain walls bounded by cosmic strings for $v_F=10^{15}\,\mathrm{GeV}$ and $v_{Z_2}=1\,\mathrm{GeV}$. The curve passes through the NANOGrav frequency bins shown for comparison \cite{NANOGrav:2023gor}.}
  \label{fig:GW-spectrum-nanograv}
\end{figure}

Because the infrared tail emerges at higher frequencies than in the pure string scenario, a non-observation at pulsar timing arrays (PTAs) does not necessarily exclude the hybrid defect model. Meanwhile, the upper bound of $v_{\rm F} \lesssim 3\times 10^{15}$ GeV set by the LVK collaboration applies equally to both pure strings and hybrid cosmic defects, as their respective gravitational wave signals remain completely identical within LVK frequencies. For different choices of the discrete $Z_2$ breaking scale at fixed $U(1)_F$ scale, the model predicts an SGWB spectrum with a characteristic $f^3$ infrared slope in the $\mu$Hz-to-Hz range. As shown in Fig.~\ref{fig:GW-spectrum-1}, for $v_F=10^{15}\,\mathrm{GeV}$ part of the parameter space lies within the projected reach of the next-generation SKA, giving a low-frequency test of the hybrid-defect scenario that is absent in the corresponding pure-string limit. Figure~\ref{fig:GW-spectrum-nanograv} further illustrates that, for the representative choice $v_F=10^{15}\,\mathrm{GeV}$ and $v_{Z_2}=1\,\mathrm{GeV}$, the hybrid-defect spectrum may possibly have implications for the signal observed in the NANOGrav measurements \cite{NANOGrav:2023gor}, providing an additional low-frequency target for PTA searches, see Fig. \ref{fig:GW-spectrum-nanograv}.


\section{Machine-learning Approach for Efficient SNR analysis}
\noindent Interferometers are instruments that measure displacements in terms of dimensionless strain-noise denoted as $h_{\text{GW}}(f)$. This strain-noise is related to the amplitude of gravitational waves (GWs) and can be converted into an energy density, expressed by the formula,

\begin{equation}
  \Omega_{\text{exp}}(f)h^2 = \frac{2\pi^2 f^2}{3\mathcal{H}_0^2} h_{\text{GW}}(f)^2h^2.
\end{equation}


\noindent Here, $\mathcal{H}_0$ represents the present-day Hubble rate ($\mathcal{H}_0 = h \times 100 \frac{\text{km/s}}{\text{Mpc}}$) and h stands for reduced Hubble constant.
To assess the likelihood of detecting the primordial GW background, we calculate the signal-to-noise ratio (SNR) using the experimental sensitivity for the noise curves $\Omega_{\text{exp}}(f)h^2$, either given for presently running experiments such as PTA and LVK or projected for future experiments such as LISA and ET. In Fig.~\ref{fig:GW-spectrum} and Fig.~\ref{fig:GW-spectrum-1}, we show the power-law integrated (PLI) sensitivity curves as discussed in~\cite{Thrane:2013oya} for various GW experiments. The color-shaded regions depict the future sensitivity prospects in each of the GW missions mentioned above. These power-law sensitivity curves are drawn based on the following assumption: the expected GW spectrum from cosmic defects can be represented in a power-law form, that is, $\Omega_{\rm GW} \sim f^b$, where $b$ is the spectral index of the frequency slope. Usually, the shaded region that falls inside such a PLI curve accounts for the range of parameters where such a power-law model GW signal will be detected with quite a high signal-to-noise ratio (SNR). These power law curves give us a useful way to represent phenomenological predictions. However, for reliable and concrete analysis and a complete understanding of our parameter space involving the hybrid defects, we next delve into our own independent direct SNR computation to assess the detectability of GW for each individual GW mission; see Ref.~\cite{Caprini:2018mtu} for further details. The SNR is determined by the following formula,

\begin{equation}\label{eq:SNR}
  \text{SNR} \equiv \sqrt{\tau \int_{f_{\text{min}}}^{f_{\text{max}}} df \left(\frac{\Omega_{\text{GW}}(f)h^2}{\Omega_{\text{exp}}(f)h^2}\right)^2}.
\end{equation}


\noindent In this analysis, we use $h = 0.7$ \cite{ParticleDataGroup:2024cfk} and an observation time of $\tau = 4$ years
for all GW detectors except the PTA and SKA based detectors for which we use $\tau = 20$ years. We use a detection threshold of SNR $\geq 10$. 
It is important to note that this formula for SNR calculation is derived under the weak signal approximation, where the GW signal is much smaller than the instrumental noise~\cite{Allen:1997ad}. While it may overestimate the true SNR for strong signals, we adopt it here for both weak and strong GW signals for simplicity purposes in this paper. 

\textit{{Astrophysical Foreground:}} Before moving on to the details of SNR, let us appreciate the issue related to the complex problem of the separation of the unavoidable astrophysical background from the 
possible cosmological background (stochastic GW background) of primordial origin that we allude to. It is still a topic under intensive and vivid investigation \cite{Caprini:2019pxz,Flauger:2020qyi,Boileau:2020rpg, Martinovic:2020hru}. To name a few such astrophysical sources, the already observed binary black hole (BH-BH) \cite{TheLIGOScientific:2016qqj, Abbott:2016nmj, TheLIGOScientific:2016pea, Song:2024pnk} and 
binary neutron star (NS-NS) by LIGO/VIRGO \cite{TheLIGOScientific:2017qsa}. This astrophysical foreground presently involves quite large uncertainties related to merger rates and astrophysical modeling, etc. Therefore, it will depend how well we will be able to resolve individual sources contributing to the GW signal in the detectors. For instance, the inspiral phase of the merger of black holes compact binaries, one expects a background of the form 
\bea \label{eq:astro}
\Omega^{\rm GW}_{\rm binaries} = \Omega_{\rm costant}\bigg(\frac{f}{25 \text{Hz}}\bigg)^{2/3} \times \theta(f_{\rm cut} - f ) \, ,
\eea 
where $\Omega_{\rm constant}$ denotes a constant value which will be extracted from forthcoming observations. Its value is expected to be approximately $\Omega_{\rm CBC} \sim 10^{-9}$~\cite{LIGOScientific:2017zlf,Boileau:2020rpg} with $f_{\rm cut} \sim 3 \times 10^3$ Hz arising from the lightest binary of the size of a solar mass. 
Now, in order to extract the cosmological signal and distinguish between the SGWB sourced by walls bounded by strings in our analysis and that generated by the astrophysical foreground, we assume that we will be able to subtract the astrophysical signals expected in the BBO, ET, or CE windows, see Refs.~\cite{Cutler:2005qq,Regimbau:2016ike} for details. Besides, in the LISA GW detector range, the dominant astrophysical sources of GW may arise from the galactic and extra-galactic binary white-dwarf foregrounds, see Refs.~\cite{Farmer:2003pa, Rosado:2011kv, Moore:2014lga} and should be subtracted \cite{Kosenko:1998mv,Adams:2010vc, Adams:2013qma}. There are steady developments in formulating methods for the subtraction of such astrophysical foregrounds from cosmological backgrounds (see for example \cite{vanRemortel:2022fkb}). However, we assume in our analysis that this astrophysical background can be exactly removed. As is evident from Eq.~\eqref{eq:astro}, typically the GW spectrum generated by the astrophysical foreground increases with frequency as $f^{2/3}$ (see Ref.~\cite{Zhu:2012xw} for further details), whose spectral shape is completely different from the GW spectrum from walls bounded by strings predicted here. This is how we envisage to pin down the GW signals from topological defects precisely.

\subsection*{Data Generation and Machine-Learning Surrogate Training}

We consider our two independent parameters to be $(v_F, v_{Z_2})$. For an SNR analysis over a wide region of the $(v_F, v_{Z_2})$ parameter-space plane, we have to evaluate the GW spectrum for a large number of $(v_F, v_{Z_2})$ pairs over each detector's frequency band.
 Computing the spectrum from Eq.~(\ref{eq:GWs2}) requires numerically solving for $t_i$. Here $t_i$ denotes the loop formation time appearing in Eq.~(\ref{eq:GWs2}); for a given observed frequency and harmonic mode it is obtained implicitly from the loop-evolution relation in Eq.~(\ref{eq:lengthLoss}) and summing a large number of terms up to $k=10^4$. We take \(k_{\max}=10^4\) because summing additional modes beyond this does not lead to any appreciable change in \(\Omega_{\rm GW} h^2\), so the harmonic sum is effectively converged at this value. This becomes computationally expensive, especially for dense scans with many $(v_F, v_{Z_2})$ pairs. Our use of machine learning is modest: we train an MLP on gravitational-wave spectra computed from the full numerical calculation and then use the trained network as a fast interpolating surrogate for repeated SNR scans over the model parameter space. Similar neural-network surrogate strategies have recently been used to emulate theoretically computed gravitational-wave spectra and other expensive gravitational-wave simulation outputs, replacing repeated numerical evaluations by fast surrogate predictions~\cite{Tian:2025zlo,Thomas:2025rje}. For training and validation, we used a final set of \(155\) cleaned spectra. The starting dataset contained \(175\) exact spectra obtained using the full numerical implementation based on the public \texttt{CosmicStringGW} code repository associated with Ref.~\cite{Fu:2024rsm}. Of these, \(171\) spectra were generated by running the full numerical calculation on the MIT SubMIT cluster~\cite{bendavid2025submitphysicsanalysisfacility} over the logarithmic grid
\(v_F \in \{10^1,\dots,10^{18}\}\,\mathrm{GeV}\) and
\(v_{Z_2}\in \{10^0,\dots,10^{17}\}\,\mathrm{GeV}\), subject to the condition \(v_F>v_{Z_2}\). The remaining four benchmark spectra, with
\(v_{Z_2}=246\,\mathrm{GeV}\) and
\(v_F=10^{12},10^{13},10^{14},10^{15}\,\mathrm{GeV}\), were taken directly from the same public code repository. After preprocessing and cleaning, \(20\) spectra were discarded because no valid points remained after applying the cleaning cuts and the floor \(\Omega_{\rm GW}h^2>10^{-30}\). This left the final set of \(155\) spectra used for training and validation.

We train the surrogate on a point-wise regression dataset with standardized inputs (mean-subtracted, variance-normalized). The surrogate model learns the mapping 
%
\begin{equation}
((v_F/{\rm GeV}),\,(v_{Z_2}/{\rm GeV}),\,(f/{\rm Hz}))
\;\longmapsto\;
(\Omega_{\rm GW}h^2)\,.
\end{equation}
Each training sample therefore corresponds to one frequency point from one exact spectrum, with three input features and one target value. For the numerical implementation we work in logarithmic variables because both the frequency and the GW amplitude span many orders of magnitude, and the input standardization is performed using the mean and variance of the training split only. We assign a spectrum ID to each spectrum to prevent leakage in the training/validation split. This is important because one exact spectrum contributes many neighboring frequency points that are highly correlated. By using the spectrum ID as the grouping variable, all points from a given spectrum are kept entirely in either the training set or the validation set within a fold, rather than being randomly split across both.
To prioritize accuracy in the detectable frequency band, we use a weighted mean absolute error (MAE) loss,
\begin{equation}
\mathcal{L}
= \Big\langle w(f)\,\big|\log_{10}(\Omega_{\rm GW}h^2)_{\rm pred}-\log_{10}(\Omega_{\rm GW}h^2)_{\rm true}\big|\Big\rangle,
\qquad
w(f)=
\begin{cases}
5, & -9 \le \log_{10}(f/{\rm Hz}) \le 5,\\
1, & \text{otherwise}.
\end{cases}
\label{eq:weighted_mae}
\end{equation}
This ensures that, while we emphasize the detectable band, the model still learns the essential spectral features from out-of-band frequencies.
We train five models using 5-fold group cross-validation, so that in each fold about 80\% of the pointwise samples are used for training and about 20\% are held out for validation. {Concretely, we use GroupKFold with 5 folds on the 155 retained spectra, so each fold holds out 31 full spectra for validation and uses the remaining 124 for training. The hybrid-spectrum surrogate itself is a fully connected MLP with 4 hidden layers of width 256, ReLU activations, LayerNorm, dropout 0.05, batch size 2048, AdamW optimizer with learning rate $10^{-3}$, and early stopping with patience 1000 epochs. For all downstream SNR calculations we use the ensemble-mean prediction of the five fold models.} In 5-fold group cross-validation, we train five MLP models with different spectra falling in the training/validation split. This helps us use the dataset efficiently and also ensures that the model is tested on hard examples, which may not end up in the validation set if we trained just one model. Finally, for the SNR calculations we use the ensemble-mean prediction of these five trained MLP models. The MLP architecture used, along with the hyperparameters and the training/validation setup, are summarized in Table~\ref{tab:surrogate_mlp}. {At the spectrum level, the corresponding validation gives a global mean MAE of 0.0309 in the selection band $(10^{-9}$--$10^{3})$ Hz after clipping $\Omega_{\rm GW}h^2\ge 10^{-18}$ for the ranking.}

\begin{table}[t]
\centering
\caption{\it Summary of the MLP architecture and training/validation setup.}

\label{tab:surrogate_mlp}
\begin{tabular}{p{0.33\linewidth} p{0.62\linewidth}}
\hline\hline
\textbf{Component} & \textbf{Specification} \\
\hline

Task & Predict $\log_{10}(\Omega_{\rm GW}h^2)$ \\
Inputs & $(\log_{10} v_F,\ \log_{10} v_{Z_2},\ \log_{10} f)$ \\
Target & $\log_{10}(\Omega_{\rm GW}h^2)$ \\
Input preprocessing & Standardize inputs (fit on training split only) \\
Model class & MLP (fully connected, feed-forward) \\
Hidden layers & 4 hidden layers, width 256 \\
Activation & ReLU \\
Normalization & LayerNorm (hidden layers) \\
Loss & Weighted L1 loss (MAE) \\
Sample weighting & $w=5$ for $-9\le \log_{10}f\le 5$, else $w=1$ \\
Optimizer & AdamW \\
Learning rate & $10^{-3}$ \\
LR schedule & Cosine annealing \\
Batch size & 2048 \\
Training stop & Max 5000 epochs; early stopping patience 1000 epochs \\
Validation & 5-fold group cross-validation (spectrum-wise holdout) \\
Ensemble & 5 models (fold ensemble); use mean prediction \\
\hline\hline
\end{tabular}
\end{table}

\begin{figure}[t]
  \centering
  \includegraphics[width=1\linewidth]{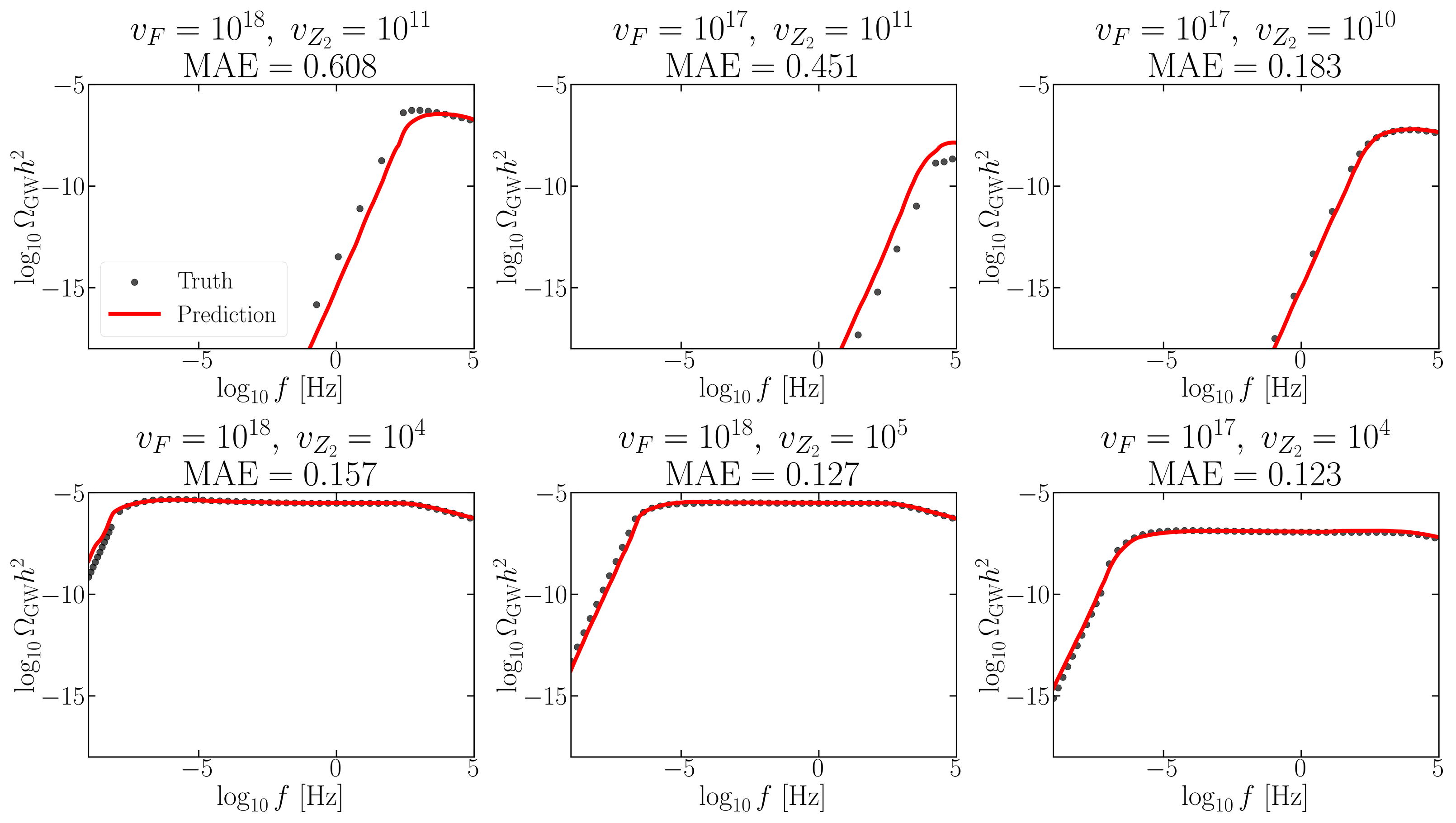}
  \caption{\it Six worst-performing spectra, ranked from highest to lowest MAE, computed over the frequency band $(10^{-9}$--$10^{3})\,\mathrm{Hz}$. Blue points denote the reference (``truth'') spectra from the numerical calculation, while the red curve shows the corresponding MLP-surrogate prediction (out-of-fold, from 5-fold cross-validation). For ranking, the MAE is evaluated after clipping the spectrum to $\Omega_{\rm GW }h^2\ge 10^{-18}$. Over the same frequency band, the global mean MAE across all spectra is $0.0309$.
}
  \label{fig:ml_worst6_hybrid}
\end{figure}

\begin{figure}[t]
  \centering
  \includegraphics[width=1\linewidth]{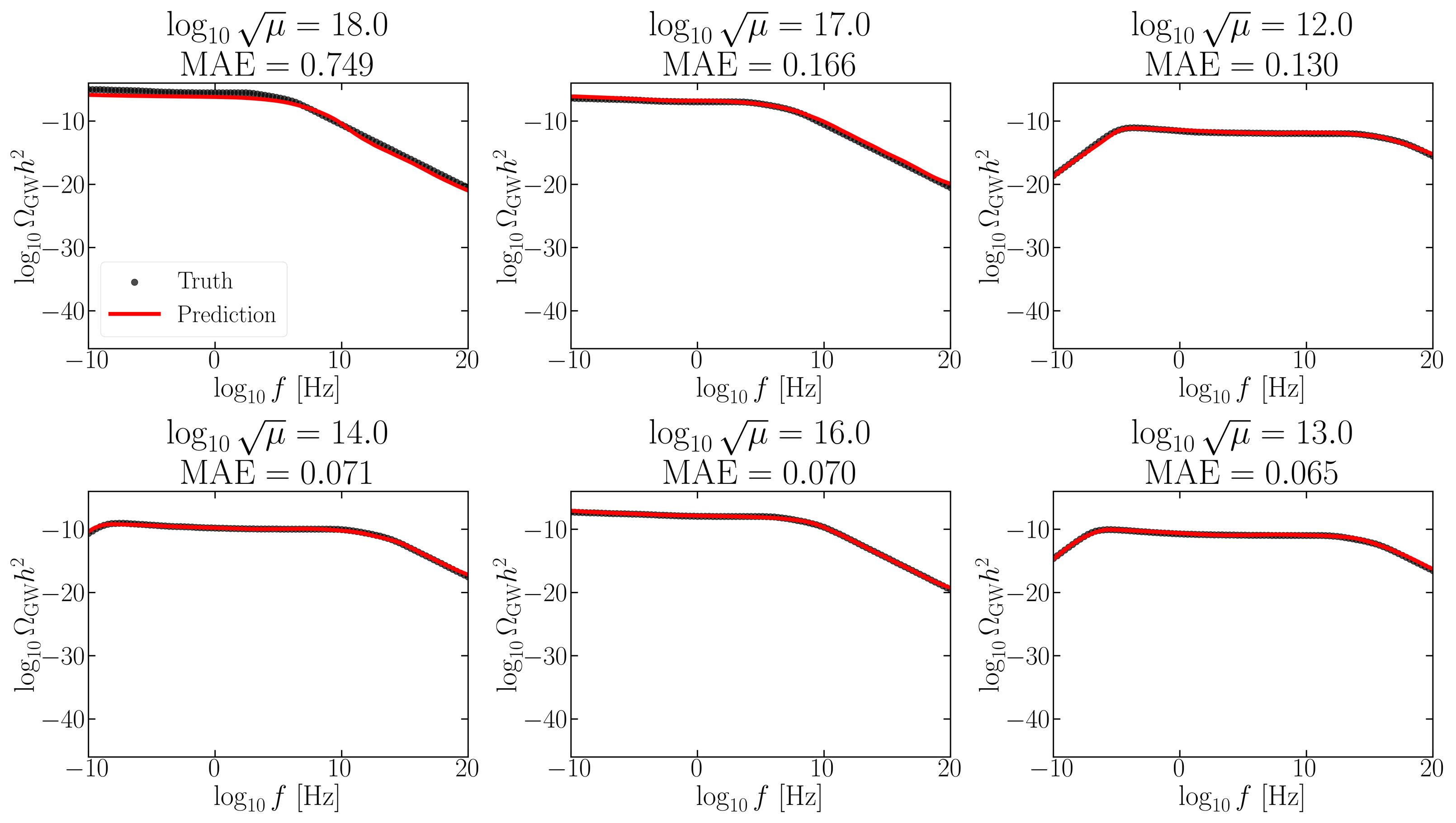}
  \caption{\it Six worst-performing cosmic-string spectra for the auxiliary surrogate used in turning-point detection. Black points denote the reference cosmic-string spectra and the red curves show the corresponding surrogate predictions (out-of-fold). The ranking MAE is computed over \(f\le 10^{3}\,\mathrm{Hz}\) after clipping \(\Omega_{\rm GW}h^2\ge 10^{-18}\). The global mean MAE over all cosmic-string spectra in the same band is \(0.0903\). Panel titles show \(\log_{10}\sqrt{\mu}\) and the per-spectrum MAE.}
  \label{fig:ml_worst6_string}
\end{figure}

Figure~\ref{fig:ml_worst6_hybrid} compares the reference (``truth'') hybrid spectra from the full numerical calculation with the corresponding MLP-surrogate predictions, using out-of-fold predictions from the 5-fold group cross-validation. The six spectra shown are the worst-performing hybrid spectra under the MAE ranking, computed over the frequency band $(10^{-9}$--$10^{3})\,\mathrm{Hz}$ after clipping the spectrum to $\Omega_{\rm GW}h^2\ge 10^{-18}$. Over the same frequency band, the global mean MAE across all hybrid spectra is $0.0309$, showing that the surrogate reproduces the numerical spectra accurately even for the largest-error validation examples. Figure~\ref{fig:ml_worst6_string} shows the analogous validation for the auxiliary cosmic-string surrogate used later in the turning-point analysis.

\begin{figure}[t]
  \centering
  \includegraphics[width=1\linewidth]{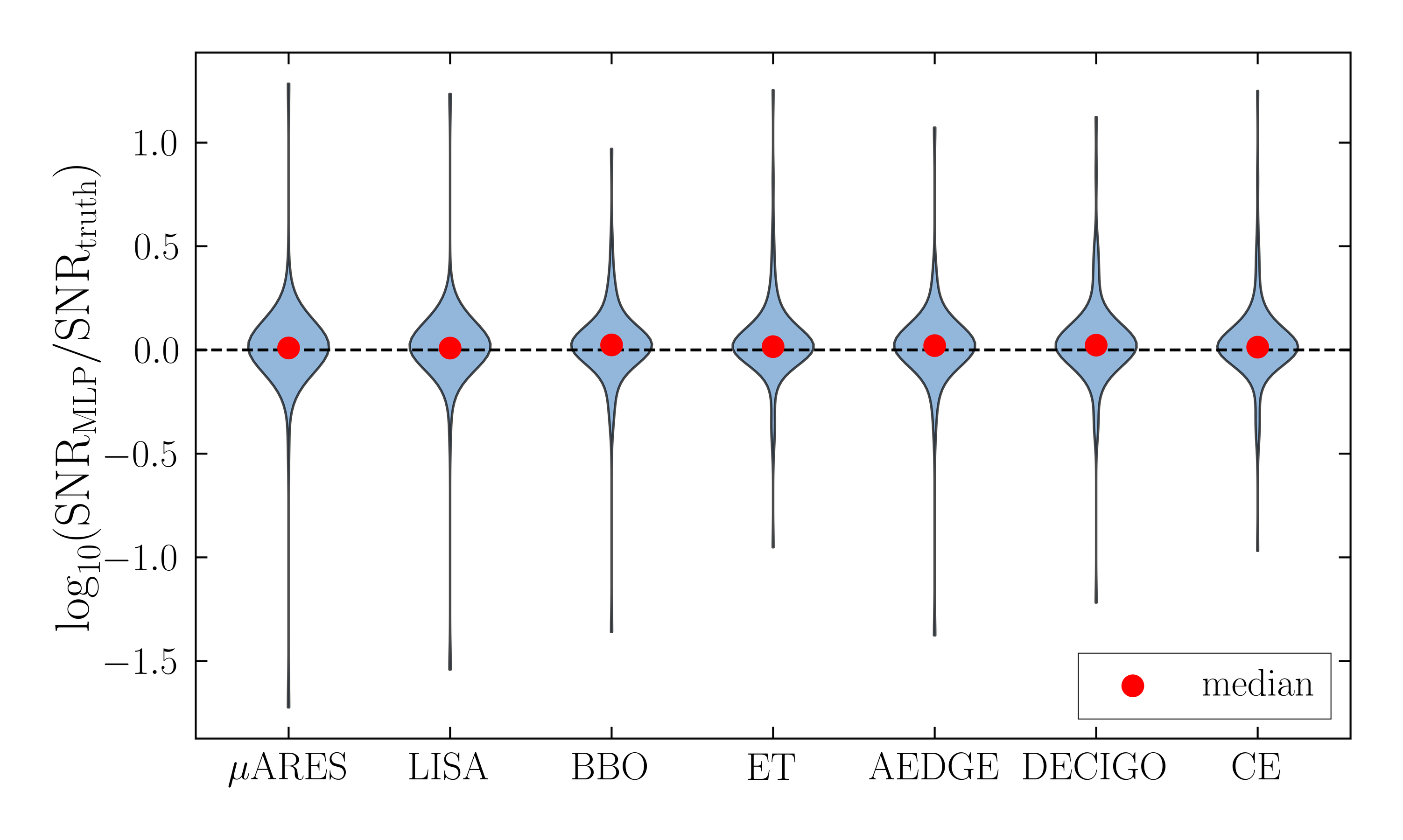}
  \caption{\it Distribution of \(\log_{10}(\mathrm{SNR}_{\rm MLP}/\mathrm{SNR}_{\rm truth})\) across detectors, evaluated on the 155 reference spectra after restricting to each detector's frequency band. Red points mark medians; the dashed horizontal line at 0 corresponds to perfect agreement. Median SNR biases are small: \(\mu\)ARES (0.010), LISA (0.009), BBO (0.024), ET (0.016), AEDGE (0.021), DECIGO (0.023), and CE (0.014), corresponding to multiplicative factors of \(1.02\)–\(1.06\). The 95th-percentile deviations remain within \(\sim 0.17\)–\(0.36\) dex (factors \(\sim 1.47\)–\(2.27\)).}
  \label{fig:snr_bias_all}
\end{figure}

To show the impact of ML on the final SNR estimates, we present the distribution of \(\log_{10}(\mathrm{SNR}_{\rm MLP}/\mathrm{SNR}_{\rm truth})\) across detectors, evaluated on the 155 reference spectra after restricting to each detector's frequency band in Fig.~\ref{fig:snr_bias_all}. We find that the median SNR biases are small: \(\mu\)ARES (0.010), LISA (0.009), BBO (0.024), ET (0.016), AEDGE (0.021), DECIGO (0.023), and CE (0.014), corresponding to multiplicative factors of \(1.02\)–\(1.06\). The 95th-percentile deviations remain within \(\sim 0.17\)–\(0.36\) dex (where one dex denotes one order of magnitude in base 10, so a deviation of $D$ dex corresponds to a multiplicative factor $10^{D}$; the factors are \(\sim 1.47\)–\(2.27\)).

These validation figures show that the MLP is not being used as a black-box replacement for the physics calculation, but rather as a fast interpolating surrogate within the physically motivated parameter space. The overlay plots demonstrate that the network reproduces both the flat high-frequency plateau and the turnover into the infrared tail for representative difficult spectra. The detector-level SNR comparison then shows that the residual spectral errors translate into only modest biases in the final observable used in the scan. This is the key practical gain: once the surrogate is trained, repeated SNR evaluations and turning-point searches can be performed rapidly while preserving the physics encoded in the full numerical spectra.

\textit{We next show how the characteristic GW spectrum from the hybrid defect may be analyzed using a possible strategy to distinguish it from other defects, such as stable cosmic strings, under some assumptions.}

\subsection*{SNR and GW Analysis Strategy}

A high SNR does not guarantee that an observed stochastic signal can be distinguished from a pure cosmic-string spectrum: at sufficiently high (by ``sufficiently high'' we mean frequencies above the turning point, namely the regime where the relative deviation $\delta(f)$ between the hybrid and pure-string spectra falls below the chosen threshold $\delta_{\rm th}$. In that regime the detector effectively sees a spectrum that is indistinguishable from the pure cosmic-string template at the chosen tolerance) frequencies the hybrid spectrum approaches the cosmic-string spectrum. To quantify whether a given detector band actually sees the hybrid spectrum portion of the signal, we define a turning-point frequency \(f_{\rm turn}\) by directly comparing the hybrid spectrum to the pure-string spectrum at fixed \(v_F\).
For this analysis, we also require a prediction for the cosmic-string spectrum. We therefore train a separate (2-input) MLP surrogate on a set of cosmic-string spectra parameterized by \((\log_{10} f,\, \log_{10}\sqrt{\mu})\), with \(\sqrt{\mu}\simeq v_F\). Figure~\ref{fig:ml_worst6_string} shows the six worst cosmic-string spectra under the spectrum-wise validation. For this purpose, we use 19 exact cosmic-string spectra, grouped by fixed $\log_{10}\sqrt{\mu}$, and train another 5-fold GroupKFold ensemble with the same 4-layer, width-256 MLP architecture. Figure~\ref{fig:ml_worst6_string} ranks the six worst held-out spectra by the MAE in the selection band $f\le 10^{3}$ Hz, after clipping $\Omega_{\rm GW}h^2\ge 10^{-18}$ for the ranking; the global mean MAE over that same band is 0.0903.

For each \((v_F,v_{Z_2})\), we evaluate both spectra on a dense frequency grid across the detector band and compute the relative deviation
\begin{equation}
\delta(f) \equiv \frac{\left|\Omega_{\rm GW}^{\rm hybrid}(f)-\Omega_{\rm GW}^{\rm string}(f)\right|}{\Omega_{\rm GW}^{\rm string}(f)}\,.
\end{equation}
We then define \(f_{\rm turn}\) as the highest frequency in the band at which \(\delta(f)\) exceeds a chosen threshold:
\begin{equation}
f_{\rm turn} \equiv \max\{f:\delta(f)>\delta_{\rm th}\}.
\end{equation}

\begin{figure}[H]
\centering
\subfloat[AEDGE\label{fig:tp_aedge}]{%
 \includegraphics[width=0.48\textwidth]{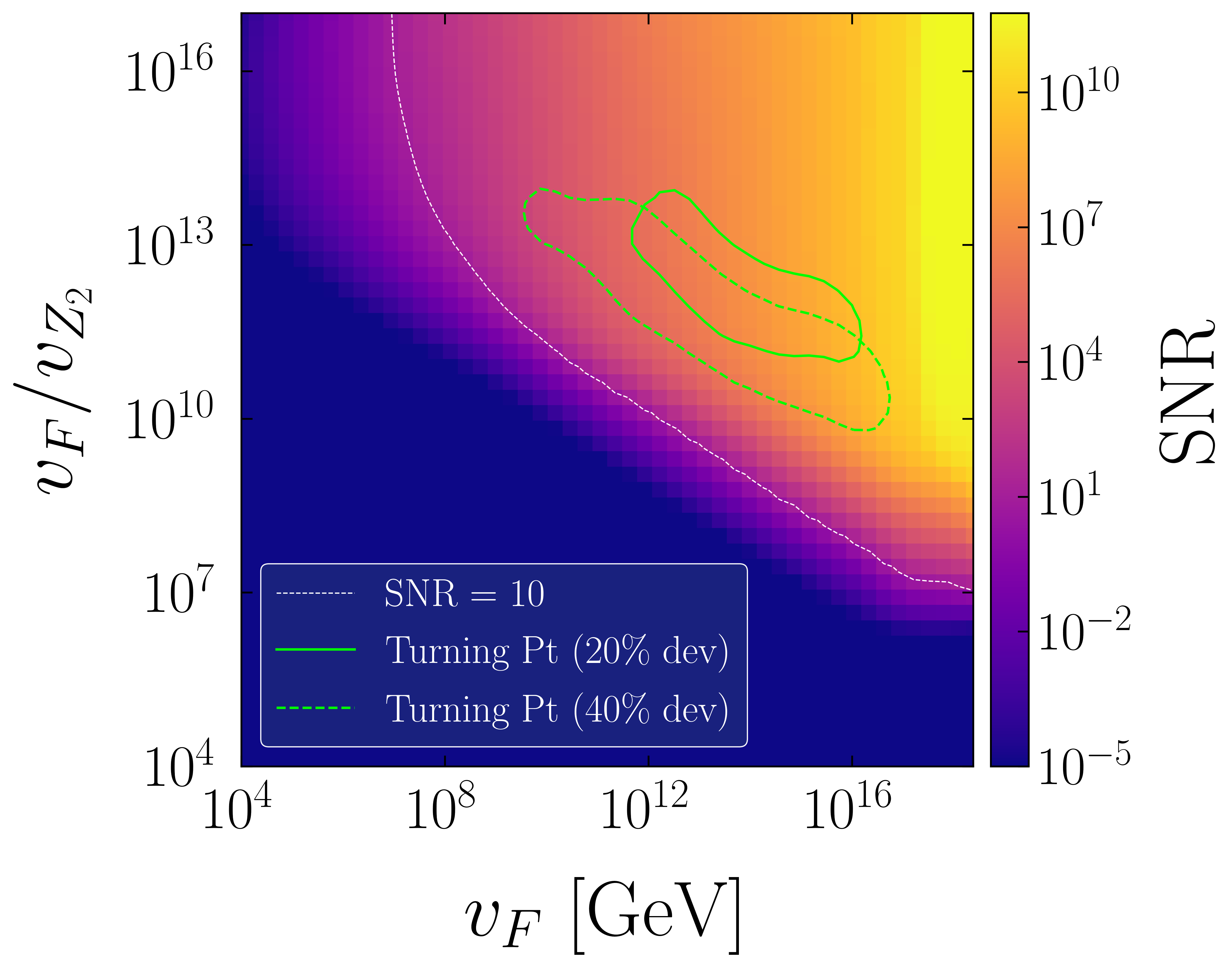}
}\hfill
\subfloat[BBO\label{fig:tp_bbo}]{%
 \includegraphics[width=0.48\textwidth]{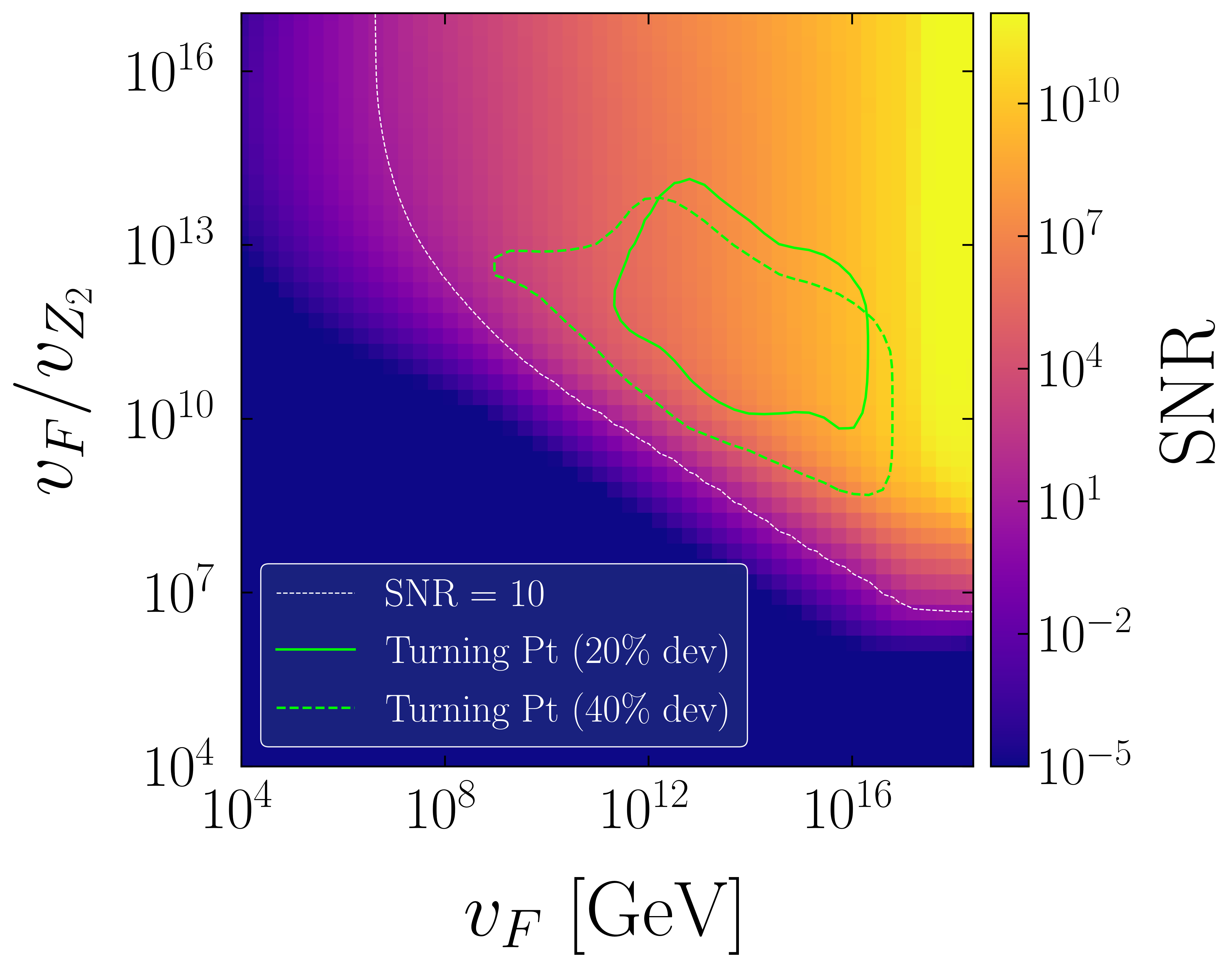}
}

\vspace{2mm}

\subfloat[Cosmic Explorer (CE)\label{fig:tp_ce}]{%
 \includegraphics[width=0.48\textwidth]{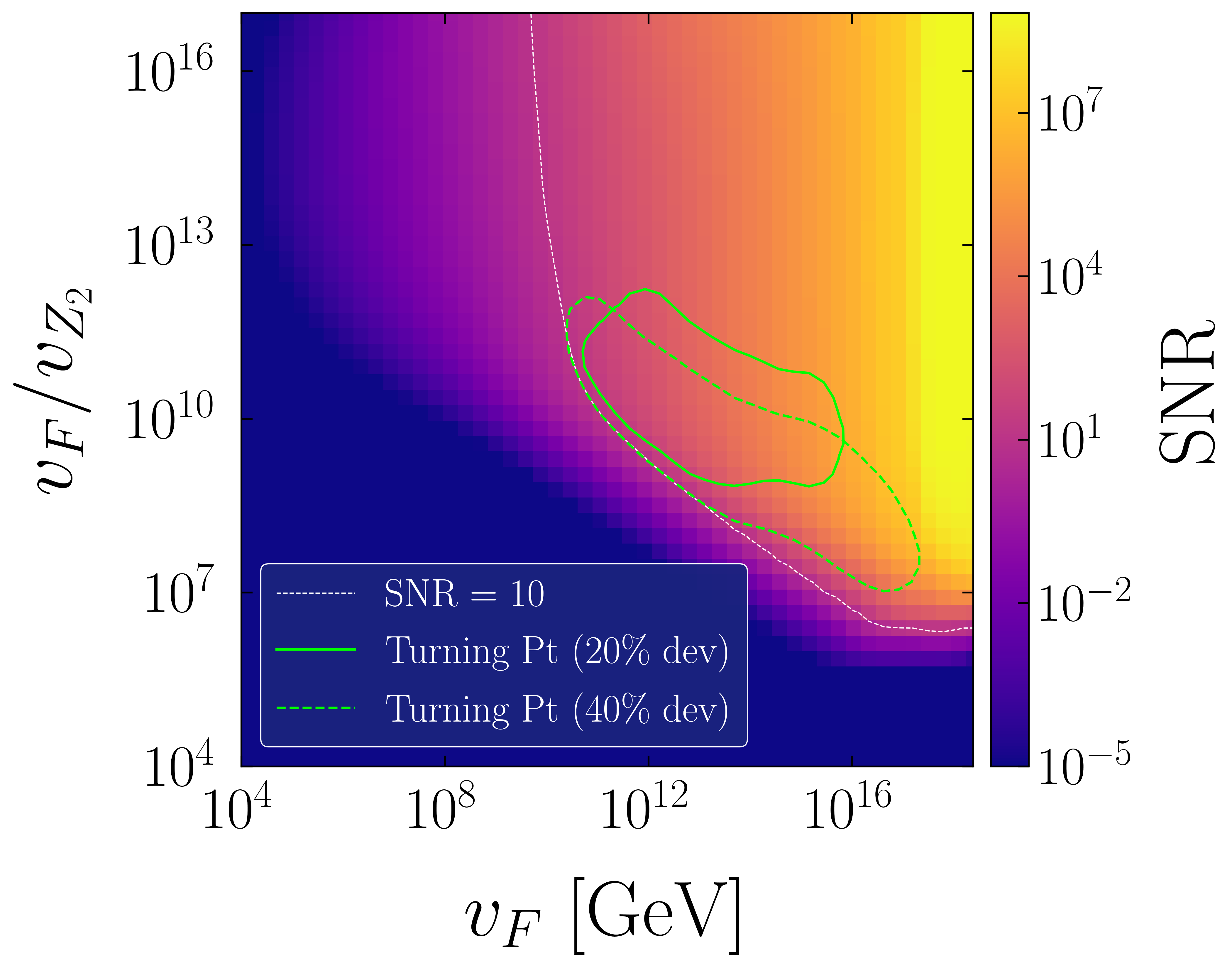}
}\hfill
\subfloat[DECIGO\label{fig:tp_decigo}]{%
 \includegraphics[width=0.48\textwidth]{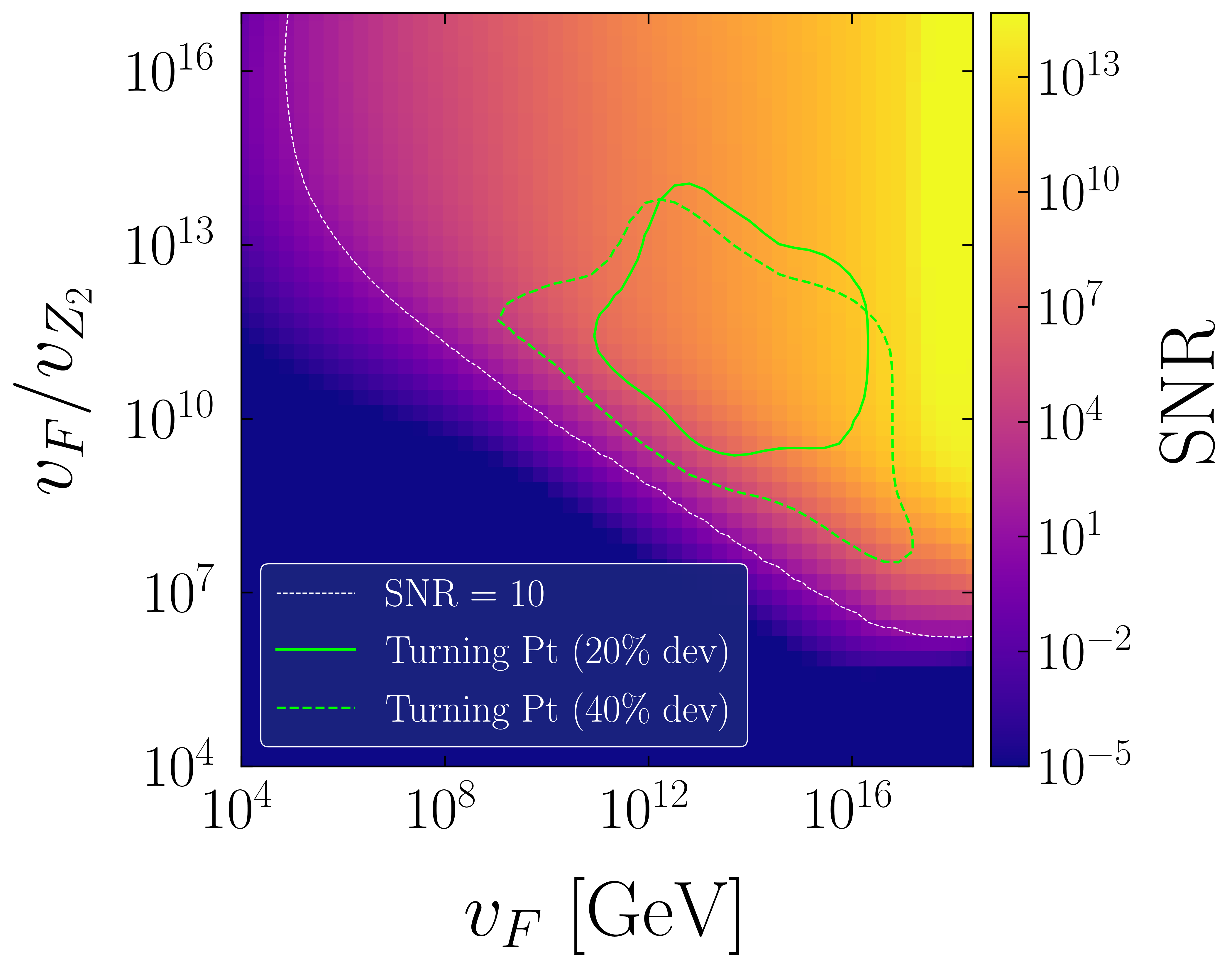}
}

\caption{\it SNR heatmaps in the \((v_F,\; v_F/v_{Z_2})\) plane for four representative detectors: AEDGE, BBO, CE and DECIGO.
The color scale shows the surrogate-predicted \(\mathrm{SNR}\) (log scale) computed using Eq.~\eqref{eq:SNR} over each detector’s frequency band.
The white dashed curve indicates the detectability threshold \(\mathrm{SNR}=10\).
Green contours indicate where the turning point lies inside the detector band, defined as the highest in-band frequency where the hybrid spectrum deviates from the pure-string template by more than \(\delta_{\rm th}\): solid for \(\delta_{\rm th}=20\%\) and dashed for \(\delta_{\rm th}=40\%\).}
\label{fig:tp_heatmaps_1}
\end{figure}

\begin{figure}[H]
\centering
\subfloat[Einstein Telescope (ET)\label{fig:tp_et}]{%
 \includegraphics[width=0.48\textwidth]{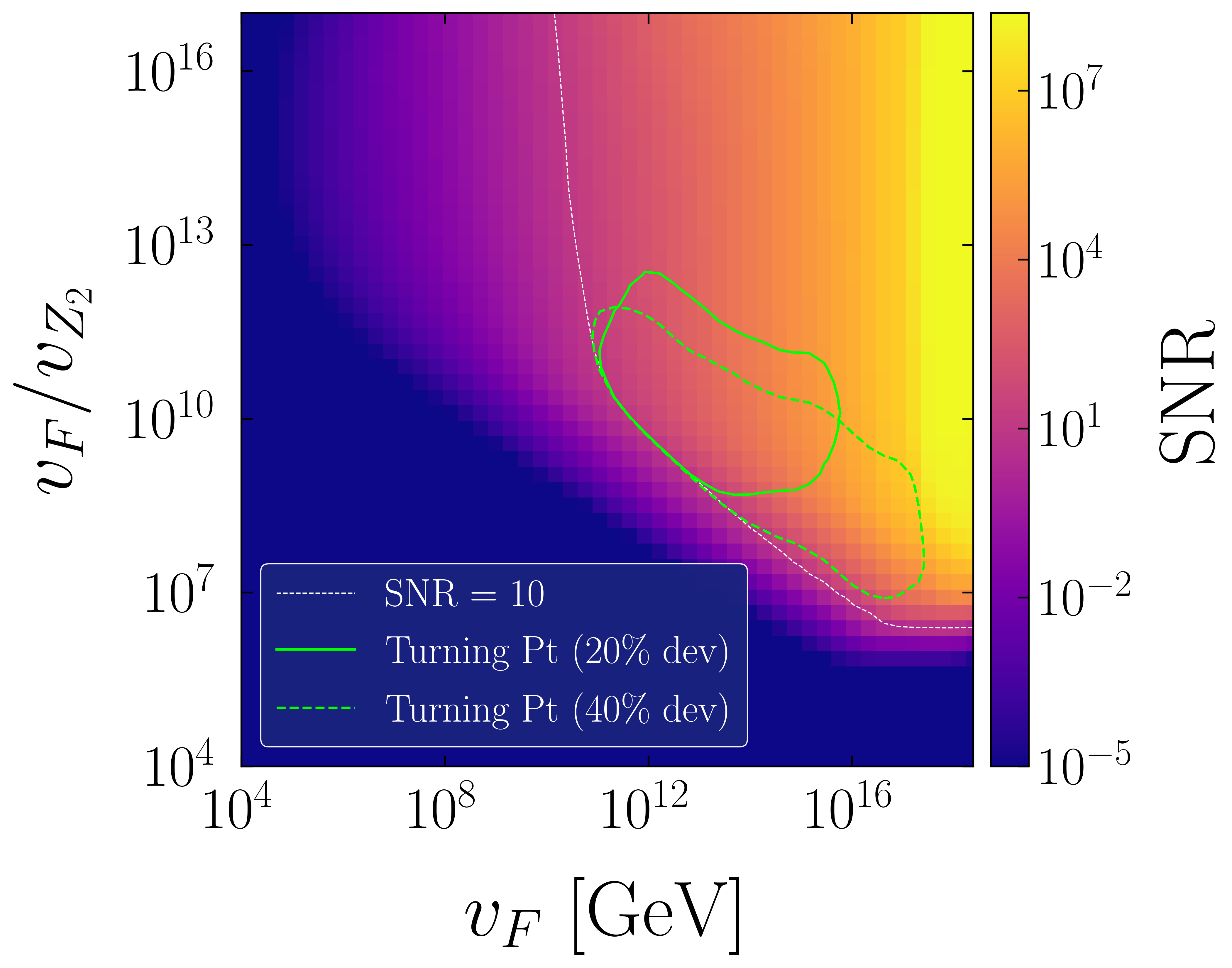}
}\hfill
\subfloat[LISA\label{fig:tp_lisa}]{%
 \includegraphics[width=0.48\textwidth]{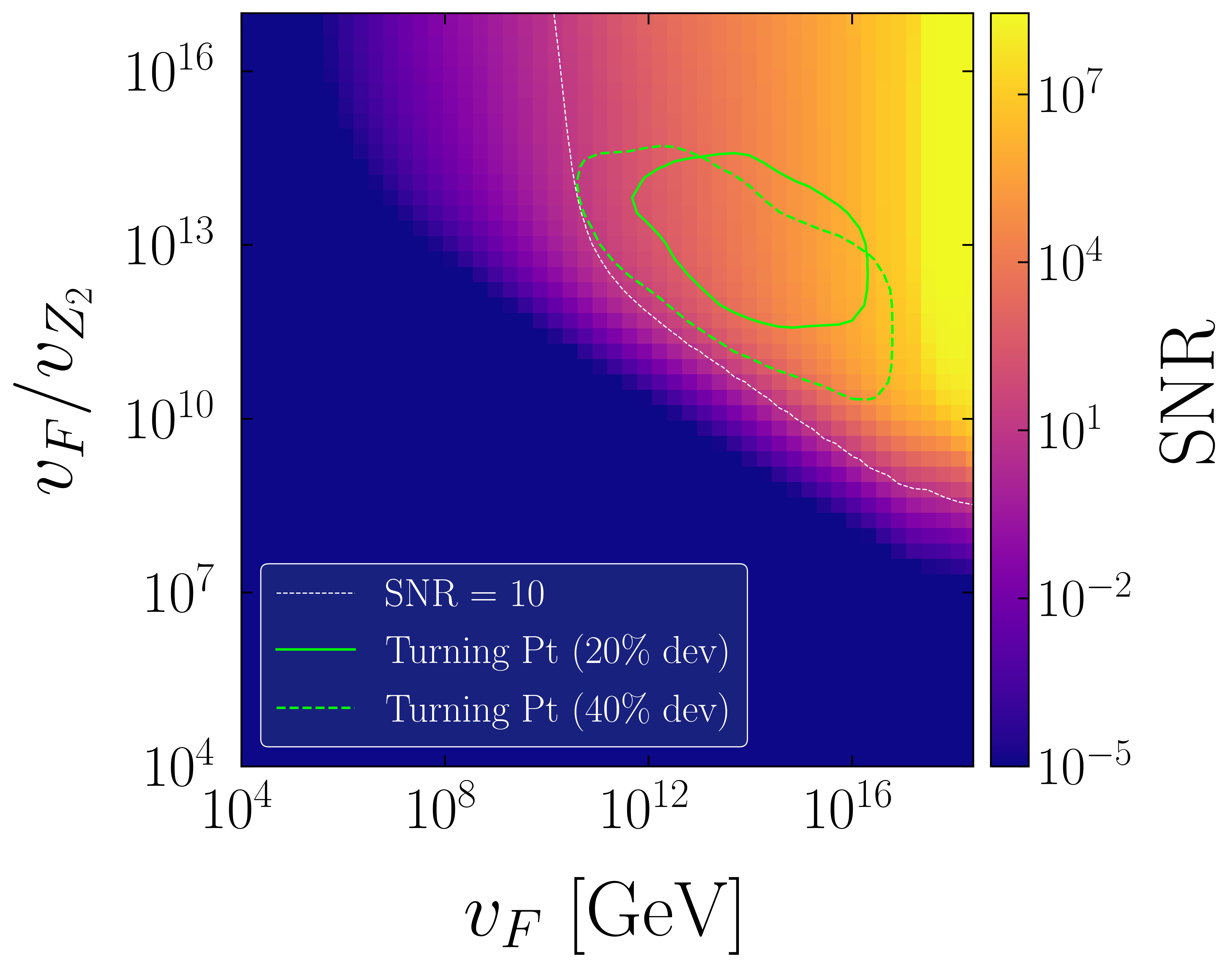}
}

\vspace{2mm}

\subfloat[\(\mu\)ARES\label{fig:tp_muares}]{%
 \includegraphics[width=0.48\textwidth]{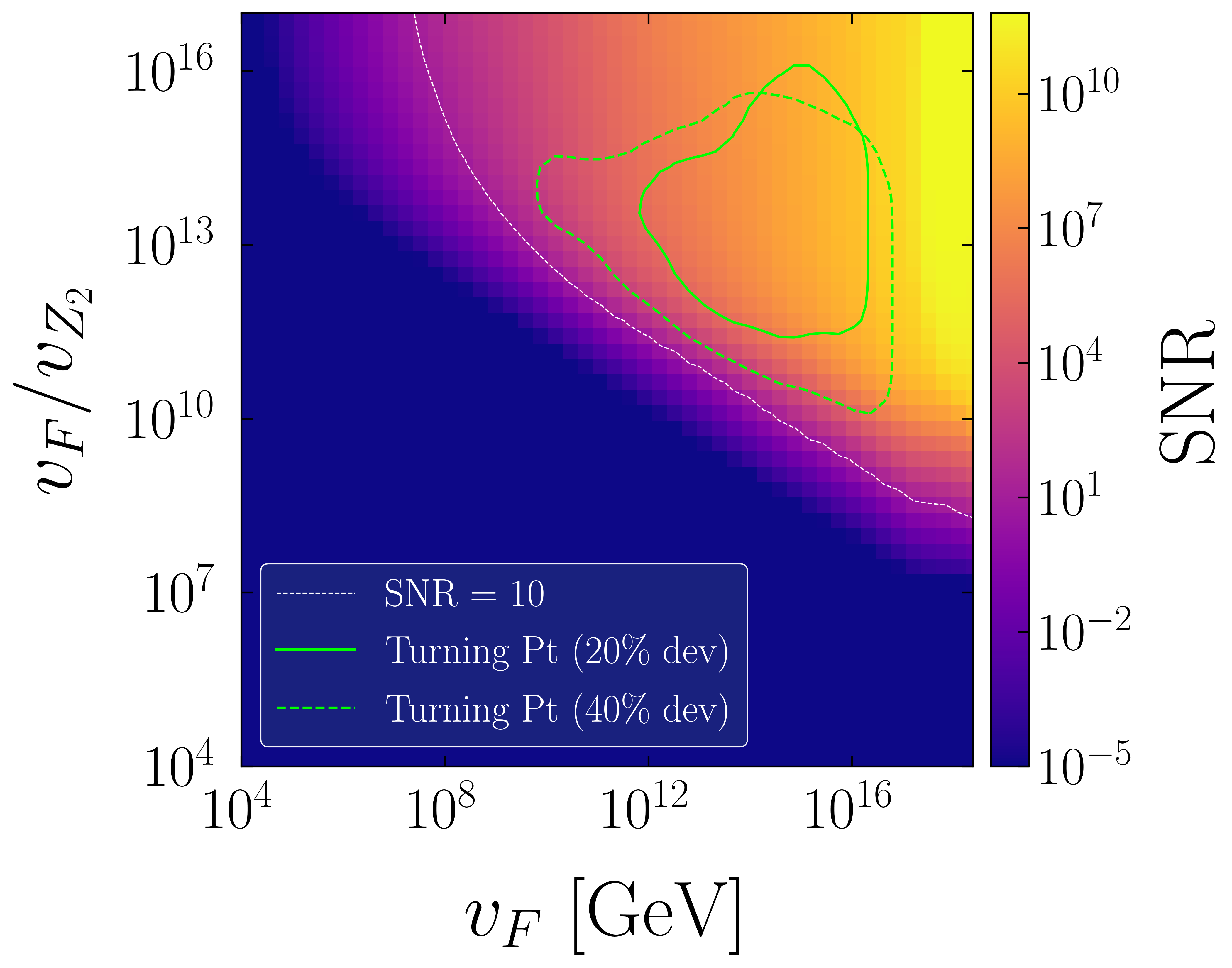}
}\hfill
\subfloat[SKA\label{fig:tp_overlay}]{%
 \includegraphics[width=0.48\textwidth]{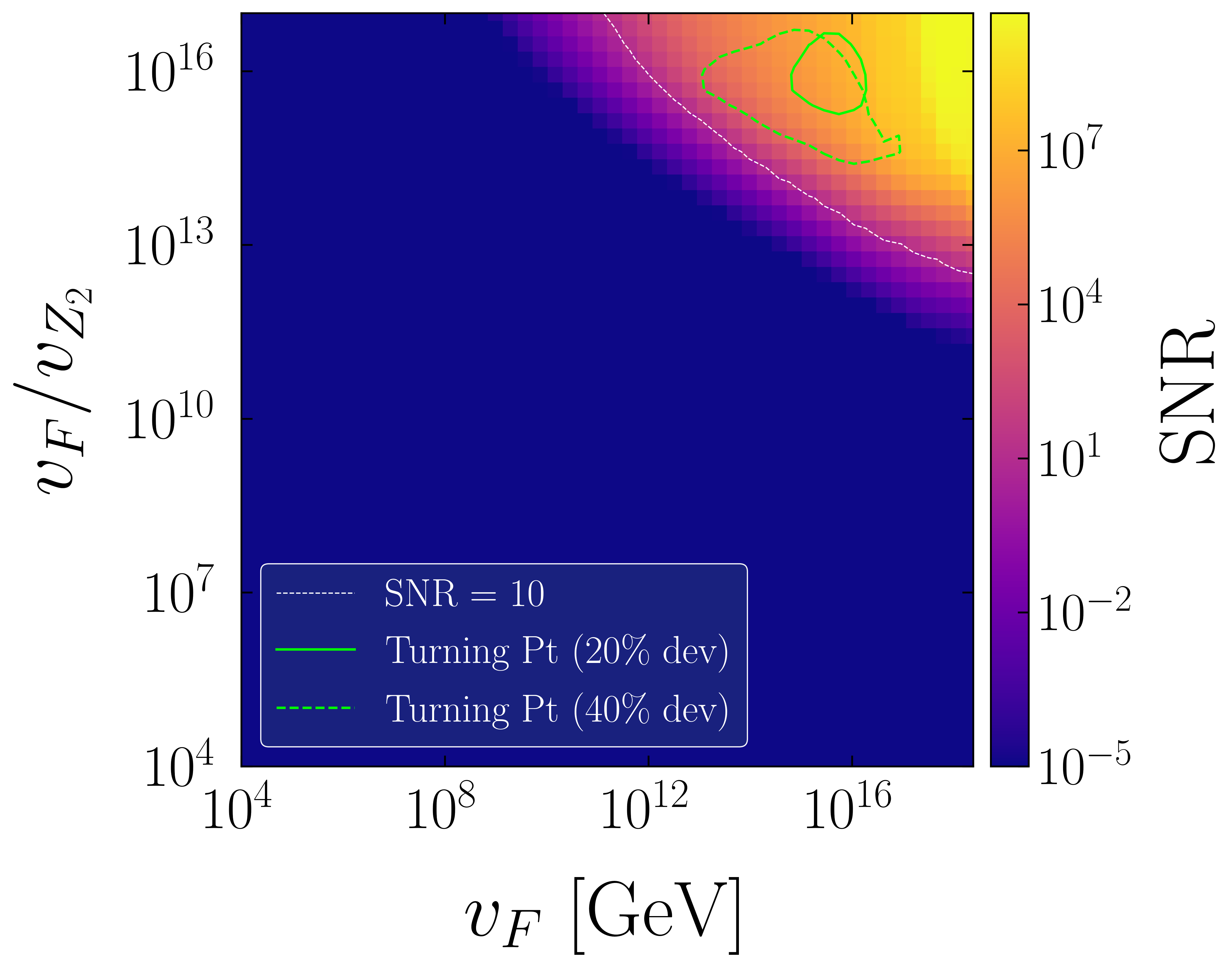}
}

\caption{\it
Same conventions as Fig.~\ref{fig:tp_heatmaps_1}.
Top/bottom-left panels show SNR heatmaps and in-band turning-point contours for ET, LISA, \(\mu\)ARES and SKA.
}
\label{fig:tp_heatmaps_2}
\end{figure}

Finally, for each detector we define a ``hybrid-detectable'' region by requiring both:
\begin{enumerate}
\item \(\mathrm{SNR}\ge 10\) computed over the full detector band, and
\item \(f_{\rm turn}\in[f_{\min},f_{\max}]\) (i.e.\ the deviation occurs within the detector band).
\end{enumerate}
In Figs.~\ref{fig:tp_heatmaps_1} and \ref{fig:tp_heatmaps_2}, the white dashed curve denotes \(\mathrm{SNR}=10\). The green contours show the subset of points where the turning-point criterion is satisfied in the same detector band, for \(\delta_{\rm th}=20\%\) (solid) and \(\delta_{\rm th}=40\%\) (dashed). The figures show the resulting SNR maps in the \((v_F,\;v_F/v_{Z_2})\) plane for each detector.
The overall trend is simple: increasing \(v_F\) boosts the signal amplitude and pushes more of the parameter space above \(\mathrm{SNR}=10\), while increasing the hierarchy \(v_F/v_{Z_2}\) delays the onset of the hybrid deviation and moves the turning point to lower frequencies.
As a result, the turning-point contours typically appear only in a subset of the GW detectable region \footnote{As usual, by detectable, we always mean SNR $> 10$.} and their location is detector-dependent, reflecting the different frequency windows.

\begin{figure}[H]
  \centering
  \includegraphics[width=0.9\linewidth]{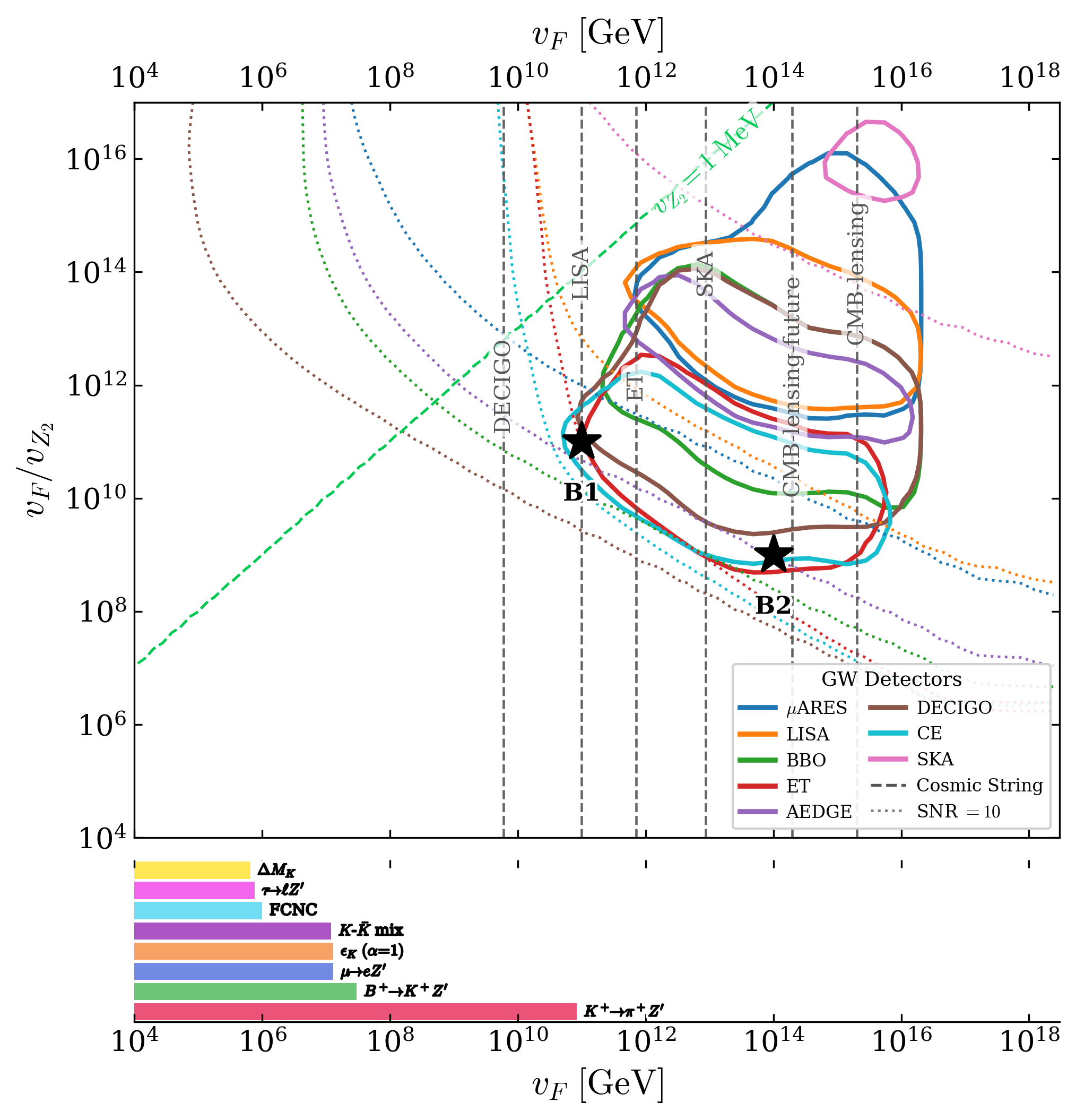}
 \caption{\it 
Scan in the \((v_F,\; v_F/v_{Z_2})\) plane (turning-point threshold \(\delta_{\rm th}=20\%\)) that captures the GW signal from the hybrid defect. Solid colored contours show where the overall \(\mathrm{SNR}\ge 10\) for the GW signal, as well as, the turning-point frequency (see text) lies within the detector. The dotted contours show the same but without the detectable turning-point frequency lying inside that particular GW detector. Green dashed diagonals mark constant \(v_{Z_2}\) reference values. The lower bar panel shows representative flavor constraints on \(v_F\). Benchmark points \(B1\) and \(B2\) are marked by stars which correspond to the GW spectrum shown in Fig.~\ref{fig:GW-spectrum} and Fig.~\ref{fig:GW-spectrum-1}. Assuming only $U(1)$ breaking, and no intermediate symmetry breaking (and therefore, no hybrid defects), the gray dashed vertical lines are shown which indicate \(\mathrm{SNR}=10\) detector reaches for the stable cosmic string case.}
\label{fig:tp_summary}
  
\end{figure}

Figure~\ref{fig:tp_summary} provides a compact view of where the hybrid signal is distinguishable from a pure cosmic-string spectrum.
Combined hybrid-detectability contours in the \((v_F,\, v_F/v_{Z_2})\) plane, using the turning-point threshold \(\delta_{\rm th}=20\%\). Solid colored closed curves denote the region where \emph{both} (i) \(\mathrm{SNR}\ge 10\) and (ii) the turning point lies inside the corresponding detector band. The dotted colored curves show the \(\mathrm{SNR}=10\) boundaries alone without requiring hybrid-detectability. Green diagonal dashed lines indicate constant \(v_{Z_2}\) reference values (\(v_{Z_2}=1~\mathrm{TeV}\) and \(1~\mathrm{MeV}\)). Gray dashed vertical lines show reference values of \(v_F\) satisfying SNR=10 for the pure cosmic-string case. Bar plots in the lower panel indicate upper bounds on the symmetry-breaking scale $v_F$ from flavor observables. Three bounds arise from rare meson decays and lepton-flavor violation in the heavy-$Z'$ regime: $K^+\to\pi^+Z'$ ($v_F < 8.3\times10^{10}$~GeV), $B^+\to K^+Z'$ ($v_F <3.0\times10^{7}$~GeV), and $\mu\to eZ'$ ($v_F < 1.3\times10^{7}$~GeV), all from Blasi et al.~\cite{Blasi:2024vew}. Additional bounds from the same reference include kaon mass splitting $\Delta M_K$ ($v_F < 6.5\times10^{5}$~GeV), CP violation in the kaon system $\epsilon_K$ with $\alpha=1$ ($v_F < 1.3\times10^{7}$~GeV), and the lepton-flavor-violating tau decay $\tau\to\ell Z'$ ($v_F < 7.6\times10^{5}$~GeV). {Future KOTO and KOTO-II experiments may possibly extend the $K^+\to\pi^+Z'$ reach up to $v_F < 5\times10^{11}$~GeV \cite{KOTO:2025uqg,KOTO:2025gvq}, thereby reaching regions where the benchmark point B1 shown in Fig.~\ref{fig:tp_summary} can possibly be tested in $\mu-$ARES and ET detectors.} We also show bounds from $K$-$\bar{K}$ mixing in the heavy-$Z'$ regime ($v_F < 1.2\times10^{7}$~GeV) from Smolkovi'{c} et al.~\cite{Smolkovic:2019jow}, and from flavor-changing neutral currents ($v_F < 1.0\times10^{6}$~GeV) from Cornella et al.~\cite{Cornella:2023zme}. {Future muon colliders may possibly reach up to $v_F < 10^{8}$~GeV from $K$-$\bar{K}$ mixing in the heavy-$Z'$ regime \cite{Cornella:2023zme}, thereby reaching regions where $SNR > 10$ can be reached in DECIGO (see Fig.~\ref{fig:tp_summary}).} Star markers {in Figs.~\ref{fig:GW-spectrum}, ~\ref{fig:GW-spectrum-1} and \ref{fig:tp_summary}} indicate two benchmark points corresponding to: \(B1:(v_F, v_F/v_{Z_2})=(10^{13}\,\mathrm{GeV},10^{11})\) and \(B2:(10^{11}\,\mathrm{GeV},10^{10})\), corresponding to \(v_{Z_2}=100~\mathrm{GeV}\) and \(10~\mathrm{GeV}\), respectively, assuming $\sim O(1)$ Wilsonian coefficients.

The green dashed line in Fig.~\ref{fig:tp_summary} denotes the $Z_2$ breaking scale $v_{Z_{2}} \sim T_{\rm DW} \geq T_{\rm BBN} \sim \mathcal{O}(1)$ MeV which is taken as an approximate limit so as to make the hybrid defect disappear before the onset of Big Bang Nucleosynthesis (BBN) which occurs roughly around $T_{\rm BBN} \sim 1 $ MeV \cite{Giudice:2000ex}.
The presence of cosmic strings leads to a static gravitational field around them, which can induce gravitational lensing and temperature anisotropies in the Cosmic Microwave Background Radiation (CMBR), marked by vertical lines as CMB lensing in Fig.~\ref{fig:tp_summary}; see Refs.~\cite{Christiansen:2010zi,Planck:2013mgr} for details.


\section{Discussion and Conclusion}\label{conclusion}
We have investigated the emergence of a hybrid topological defect, namely ``domain walls bounded by cosmic strings'', and the unique characteristic gravitational wave signal that emerges from it in the context of flavor symmetry breaking with a CP-odd DM candidate. 
We study a flavor model accommodating a DM candidate and, a hybrid defect involving walls-bounded-by-strings due to two-stage symmetry breaking. This provides a compelling test of flavor physics via GW from hybrid defects, which have a characteristic $f^3$ scaling in the infrared compared to standard $U(1)$ strings (see Fig.~\ref{fig:GW-spectrum} and Fig.~\ref{fig:GW-spectrum-1}). Because the DM candidate is a WIMP-like pseudo-scalar, it evades direct-detection constraints due to vanishingly small momentum-transfer. We also show some possible complementarity tests between GW and laboratory searches (see Fig.~\ref{fig:tp_summary}). The phenomenological motivation for the $U(1)_F$ extension stems from the necessity to incorporate fermion mass and mixing in the Standard Model (SM) of particle physics. 

An interesting outcome of the $U(1)_F$ charge assignment, motivated by the spontaneous generation of fermion masses, is that $U(1)_F$ is broken to an intermediate $Z_2$ symmetry, which is further broken around the intermediate scale $v_{Z_2}$ by the $S$ flavon field. Such a symmetry-breaking chain triggers a hybrid topological defect of ``domain walls bounded by cosmic strings'', created by the domain walls of $Z_2$ breaking being encircled by pre-existing cosmic-string loops originating from the earlier $U(1)_F$ breaking. Thus the symmetry is broken spontaneously in two steps, $U(1)_F\to Z_2\to { \mathcal{I}}$, by the VEVs of the flavon fields $\Phi$ and $S$, respectively, the last breaking playing a role in DM physics. 
The $U(1)_F$ acts exclusively on all three generations of matter, allowing for large lepton mixing. In addition to explaining the mass hierarchies of the SM fermions, it realizes the cosmological formation of hybrid defects whose GW spectrum is fully testable by ongoing and future GW detectors.

 We have studied the cosmological footprint of this model in terms of the decay of the hybrid network of walls bounded by strings into gravitational waves. The defect network evolves by losing energy via gravitational wave emission, where the observable frequency of the signal is inversely proportional to the time of evolution. The decay of the cosmic string loops created after $U(1)_F$ breaking yields a flat spectrum of gravitational waves at high frequencies. The additional tension exerted by the bounded domain walls arising after $Z_2$ breaking causes an accelerated decay of the loops, resulting in a characteristic $f^{3}$ slope of the signal at lower frequencies. The pivot frequency where this spectral change occurs is determined by the ratio of the two symmetry-breaking scales, while the high-frequency flat amplitude is representative of the higher scale (see Figs.~\ref{fig:GW-spectrum} and \ref{fig:GW-spectrum-1}). This leads to a prediction for the $f^{3}$ behaviour to be observed in the microHz to Hz band with a flat spectrum in the ultraviolet, when the $U(1)_F$ breaking scale is between $10^{12}$ GeV and $10^{15}$ GeV. In this sense, our study provides a way to test the model in upcoming gravitational wave interferometers such as LISA and ET, complementing flavor observables involving flavor non-universal $Z^{\prime}$ with $v_F$ between $10^4$ GeV and $10^7$ GeV (see Fig.~\ref{fig:tp_summary}). The IR tail of the GW spectrum appearing at relatively higher frequencies compared to the pure string case implies that non-observation at pulsar timing arrays (PTA) does not rule out signals from the hybrid defect for $v_F \geq 10^{14} $ GeV. For different choices of the discrete $Z_2$ breaking scale at fixed $U(1)_F$ scale, the model predicts an SGWB spectrum with a characteristic $f^3$ infrared slope in the $\mu$Hz-to-Hz range. As shown in Fig.~\ref{fig:GW-spectrum-1}, for $v_F=10^{15}\,\mathrm{GeV}$ part of the parameter space lies within the projected reach of the next-generation SKA, giving a low-frequency test of the hybrid-defect scenario that is absent in the corresponding pure-string limit. 
 A more complex breaking chain (e.g. $SU(N) \rightarrow U(1) \rightarrow Z_N \rightarrow Z_2 \rightarrow I$) could enhance this "non-detectability" effect; however, such an analysis is beyond the scope of the current paper. Due to $\epsilon \sim \frac{\langle\Phi\rangle}{\Lambda} \sim \frac{v_F}{\sqrt{2} \Lambda}$, with $v_F$ controlling the amplitude of the GW signal, one finds a novel correlation between PMNS and CKM structure and the observable GW spectrum that also depends on the choice of $M_{R_{ij}}$.

We developed a strategy such that the characteristic GW spectrum from the hybrid defect may be analyzed to distinguish it from other defects, such as stable cosmic strings, via prescribing a turning-point frequency \(f_{\rm turn}\) which allows to directly compare the hybrid spectrum to the pure-string spectrum at fixed $U(1)_{F}$ flavor breaking scale \(v_F\). In order to employ this strategy, we complement the exact calculation with a \textit{machine-learning surrogate} based on a multilayer perceptron (MLP), trained on spectra obtained from the full numerical treatment and then used for rapid inference in the SNR computation. This allows us to efficiently map the detectability of the hybrid signal and quantify the region in which it can be distinguished from the corresponding pure cosmic-string spectrum.
For the \(175\) sampled \((v_F,v_{Z_2})\) spectra used in this work, the numerical computation code requires a noticeable amount of cluster time. Since the jobs were run with up to \(20\) spectra evaluated in parallel on the MIT SubMIT cluster, this corresponds to about \(9\) rounds of calculations. If each round takes about \(30\)–\(60\) minutes, the full scan requires roughly \(4.5\)–\(9\) hours of wall-clock time. By contrast, once the MLP surrogate is trained, the same scan can be carried out much more quickly through simple model evaluation. This is why the \textit{machine-learning approach} is useful here: it makes repeated SNR studies over the full parameter space practical (see Figs.~\ref{fig:ml_worst6_hybrid}, \ref{fig:ml_worst6_string}, and \ref{fig:snr_bias_all}).

 In summary, this work introduces a novel direction in flavor model building that not only addresses the renowned \textit{flavor puzzle}, but also predicts detectable GW spectrum which promotes future investigations into the paradigm of \textit{complementarity between Gravitational Waves and Flavor Observables}. Gravitational wave astronomy aspires to achieve precisions that are orders of magnitude
better than the current detectors. The new era of GW detectors planned worldwide will make the dream of testing fundamental BSM mechanisms, e.g. for flavor physics, a reality in the near future.


\section*{Acknowledgement}

We thank K.S. Babu, Kenji Nishiwaki, V. Knapp-Perez, C. Moffett-Smith and Newton Nath for useful discussions.
This work made use of resources provided by subMIT at MIT Physics. A.G. acknowledges the support from the Royal Society, UK, Funding Reference: NIF\ R1\ 253963.

\bibliographystyle{apsrev4-1}

\bibliography{ref}


\end{document}